\documentclass[%
 reprint,
 amsmath,amssymb,
 aps,
]{revtex4-2}

\usepackage{dcolumn}
\usepackage{bm}
\usepackage{times}
\usepackage{amsfonts}
\usepackage{amsmath}
\usepackage{amssymb}
\usepackage[pdftex]{graphicx}
\usepackage{MnSymbol}
\usepackage{multirow}
\usepackage{hyperref}
\usepackage{braket}
\usepackage{color}
\usepackage{tikz}
\usepackage{textcomp}
\usepackage{sidecap}
\newcommand{\EOiOj}[2]{\braket{\hat{O}_{#1}\hat{O}_{#2}}}
\newcommand{\EOij}[2]{\braket{\hat{O}_{#1}^{#2}}}

\newcommand{\Yb}{$^{171}\rm{Yb}^+$}
\newcommand{\Ba}{$^{138}\rm{Ba}^+$}
\newcommand{\Unit}[1]{\,\mathrm{#1}}

\begin{document}

\preprint{APS/123-QED}

\title{Significant-loophole-free test of Kochen-Specker contextuality using two species of atomic-ions}

\author{Pengfei Wang,$^{1,2,\ast}$ Junhua Zhang,$^{3,\ast}$ Chun-Yang Luan,$^{1}$ Mark Um,$^{1}$ Ye Wang,$^{4}$\\
	Mu Qiao,$^{1}$ Tian Xie,$^{5}$ Jing-Ning Zhang,$^{2}$ Ad\'an Cabello,$^{6,7,\dagger}$ Kihwan Kim,$^{1,2,8\ddagger}$\\
\normalsize{$^{1}$State Key Laboratory of Low Dimensional Quantum Physics, Department of Physics, }\\
\normalsize{Tsinghua University, Beijing 100084, People's Republic of China }\\
\normalsize{$^{2}$Beijing Academy of Quantum Information Sciences, Beijing 100193, People's Republic of China}\\
\normalsize{$^{3}$Shenzhen Institute for Quantum Science and Engineering, }\\
\normalsize{Southern University of Science and Technology, Shenzhen 518055,  People's Republic of China}\\
\normalsize{$^{4}$Department of Electrical and Computer Engineering, Duke University, Durham, NC 27708, USA}\\
\normalsize{$^{5}$Kavli Nanoscience Institute and Thomas J. Watson, Sr., Laboratory of Applied Physics, }\\
\normalsize{California Institute of Technology, Pasadena, California 91125, USA}\\
\normalsize{$^{6}$Departamento de F\'isica Aplicada II, Universidad de Sevilla, E-41012 Sevilla, Spain}\\
\normalsize{$^{7}$Instituto Carlos~I de F\'{\i}sica Te\'orica y Computacional, Universidad de
	Sevilla, E-41012 Sevilla, Spain}\\
\normalsize{$^{8}$Frontier Science Center for Quantum Information, Beijing 100084, People's Republic of China}\\
\normalsize{$^\ast$ These two authors contributed equally}\\
\normalsize{$^\dagger$ adan@us.es}\\
\normalsize{$^\ddagger$ kimkihwan@mail.tsinghua.edu.cn}
}

\begin{abstract}
Quantum measurements cannot be thought of as revealing preexisting results, even when they do not disturb any other measurement in the same trial. This feature is called contextuality and is crucial for the quantum advantage in computing. Here, we report the first observation of quantum contextuality simultaneously free of the detection, sharpness and compatibility loopholes. The detection and sharpness loopholes are closed by adopting a hybrid two-ion system and highly efficient fluorescence measurements offering a detection efficiency of $100\%$ and a measurement repeatability $>98\%$. The compatibility loophole is closed by targeting correlations between observables for two different ions in a Paul trap, a $^{171}\mathrm{Yb}^{+}$ ion and a $^{138}\mathrm{Ba}^{+}$ ion, chosen so measurements on each ion use different operation laser wavelengths, fluorescence wavelengths, and detectors. The experimental results show a violation of the bound for the most adversarial noncontextual models and open a new way to certify quantum systems.
\end{abstract}

\maketitle

\section{Introduction}
In everyday life, whenever the measurements of two observables $A$ and $B$ yield the same values ($a$ for $A$ and $b$ for $B$) when the measurements are repeated in any order, we attribute it to the measured system possessing preexisting values revealed by every measurement and which persist after the measurements. However, this assumption fails in quantum mechanics. Quantum systems can produce correlations \cite{Klyachko2008,Cabello2008exp} between measurements which do not disturb each other and yield the same result when repeated and which, however, cannot be explained by models based on the assumption of outcome noncontextuality that states that the result is the same no matter which other compatible observables are measured in the same trial. This phenomenon, called Kochen-Specker contextuality or contextuality for sharp measurements is rooted in the Bell-Kochen-Specker theorem \cite{Bell1966,Kochen1967} of impossibility of hidden variables in quantum mechanics and is behind the power of quantum computers to outperform classical computers \cite{AB09,Raussendorf13,Howard2014,DGBR15,Bravyi18}.

Contextual correlations between sequential measurements have been observed in experiments with photons \cite{MWZ00,HLZPG03,Amselem&Cabello2009,Lapkiewicz2011,ZWDCLHYD12,DHANBSC13,AACB13,MANCB14}
neutrons \cite{YLB03}, ions \cite{Kirchmair&Cabello&Blatt2009,ZhangXiang2013,CabelloJPhome2018},
molecular nuclear spins \cite{MRCL10}, superconducting systems \cite{Jerger2016}, and nuclear spins \cite{VCTH19}. However, these experiments have ``loopholes'', as noncontextual models assisted by mechanisms that exploit the experimental imperfections can simulate the observed correlations.

Three main loopholes have been considered. The sharpness loophole follows from the observation \cite{Spekkens2005,Spekkens2014}
that the assumption of outcome noncontextuality, on which the bound of the noncontextuality inequalities is derived \cite{Klyachko2008,Cabello2008exp}, can only be justified for the case of sharp measurements, defined \cite{YuanXiao2014,Cabello2019Quantumcorrelations} as those that yield the same result when repeated and do not disturb compatible (i.e., jointly measurable \cite{Heinosaari2016}) observables.
The detection loophole \cite{Pearle1970,GargMermin1987}
exploits the lack of perfect detection efficiency and is common to Bell inequality experiments \cite{HBD15,GVW15,SMC15,RBG17}
The compatibility loophole \cite{Guhne&Cabello2010,Szangolies2013Tests,Cabello2011qutrit,HuXiaomin2016}
exploits that, in experiments with sequential measurements on the same system, the assumption that the measured observables are compatible cannot be verified.

Loophole-free Bell inequality tests \cite{HBD15,GVW15,SMC15,RBG17}
can be thought as contextuality tests that simultaneously close the detection and compatibility loopholes. However, as tests of noncontextual models, they leave open the sharpness loophole. On the other hand, there are contextuality tests free of the detection loophole and whose correlations cannot be produced by specific mechanisms exploiting the lack of perfect repeatability \cite{Kirchmair&Cabello&Blatt2009,Malinowski&HomeJP&Cabello2018}.
However, they suffer from the compatibility loophole, as they require sequential measurements performed on the same system. Therefore, a pending challenge is closing all three loopholes simultaneously in the same experiment.

For this aim, we choose a composite system of two different ions \cite{home2013quantum,Tan&Wineland2015,Ballance2015,Inlek2017,Home2018,MIT2019}, one $^{171}\mathrm{Yb}^{+}$
ion and one $^{138}\mathrm{Ba}^{+}$ ion. This dual-species system allows us performing sequential repeatable highly efficient single-shot fluorescence measurements on each of the ions. In the system, the detection loophole is naturally addressed due to the detection efficiency of 100$\%$, that is, no missing results in all trials of the experiments. We note that the detection fidelity, the probability of obtaining a correct result from a measurement, is not necessary to be perfect to close the detection loophole.

To close the compatibility loophole, we target a ``Bell-like'' \cite{MWZ00,HLZPG03}
noncontextuality inequality in which only two observables are measured per context, and each of the observables is defined on a different ion. Therefore, these two observables are ``trivially compatible ({\em i.e.}, they are simultaneously measurable in an uncontentious sense)'' \cite{Isham2001}.
In addition, the compatibility is enforced by choosing ions of different species requiring different operation laser wavelengths, fluorescence wavelengths and detectors.
\begin{figure}[ht]
	\centering
	\includegraphics[width=0.45\textwidth]{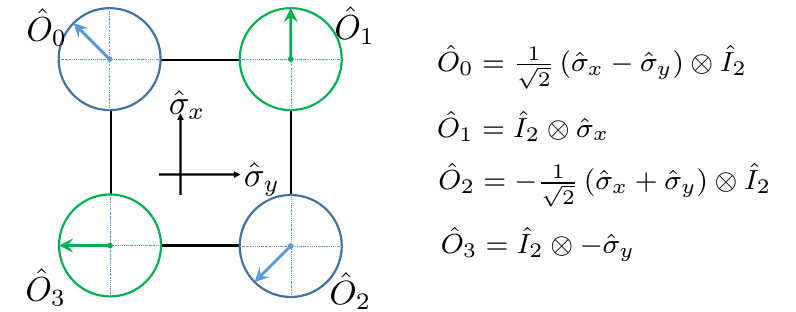}
	\caption{\textbf{The four observables and their compatibility relations.} $\hat{O}_0$ and $\hat{O}_2$ are measured on the \Yb\, ion, and $\hat{O}_1$ and $\hat{O}_3$ are measured on the \Ba\, ion. Connected observables are compatible (jointly measurable). Here, $\hat{\sigma}_x$, $\hat{\sigma}_y$ are Pauli operators and $\hat{I}_2$ is the identity operator.}
	\label{figobs}
\end{figure}

\section{Results}
The noncontextuality inequality we focus on is the only tight (i.e., strictly separating noncontextual from contextual correlations) noncontextuality inequality in the 4-cycle contextuality scenario \cite{Araujo&Cabello2013} shown in Fig.~\ref{figobs}. This is the scenario involving the smallest number of measurements that allows for contextuality for sharp measurements, as follows from a theorem by Vorob'yev \cite{Vorob'yev62,Vorob'yev63}.
This noncontextuality inequality is algebraically identical to the Clauser-Horne-Shimony-Holt (CHSH) Bell inequality \cite{Clauser1969}. It can be written as
\begin{equation}
\begin{split}
\mathcal{C}=&\EOiOj{0}{1}+\EOiOj{1}{2}+\EOiOj{2}{3}-\EOiOj{3}{0} \leq2,\\
\end{split}
\label{ineq}
\end{equation}
where each of the four observables $\hat{O}_i$ has possible results either $-1$ or $+1$, and $\EOiOj{i}{j}$ denotes the mean value of the product of the results of $\hat{O}_i$ and $\hat{O}_j$.
Unlike the CHSH Bell inequality, testing inequality (\ref{ineq}) neither require spacelike separation \cite{kleinmann2012optimal} nor adscribing the observables to two parties. Instead, it requires the measurements to be sharp.
\begin{figure}[htp]
	\centering
	\includegraphics[width=0.35\textwidth]{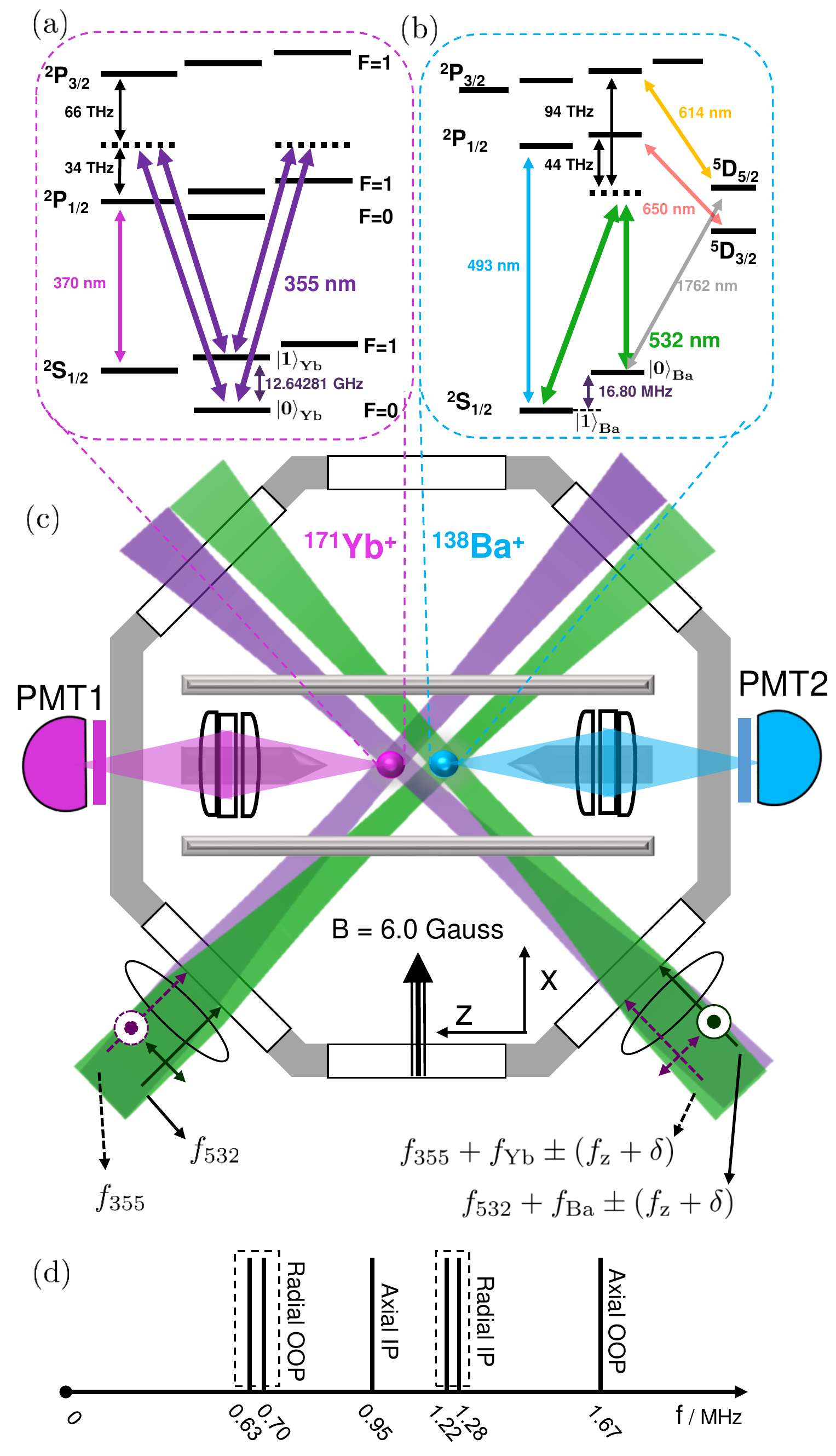}
	\caption{\textbf{Experimental setup.} (a) and (b) are the energy level diagrams of \Yb\ and \Ba\ ion, respectively. Only relevant Raman transitions are shown here. (c) Ion trap in the octagon chamber and schematic diagram for Raman beams. Two different photomultiplier tubes (PMTs) with different spectral response and filters are used to detect two ions fluorescence independently, which are located at the top and the bottom of the chamber in the actual experimental system. Solid and dashed arrows indicate the directions and the polarizations of 532 nm and 355 nm laser beams, respectively. In the figure, $f_{\textrm{Yb}}$ and $f_{\textrm{Ba}}$ are the qubit frequencies of \Yb\ and \Ba, respectively, $f_{\text{z}}=1.67$ MHz is the frequency of the axial out-of-phase (OOP) mode, and $\delta$ is the detuning of the laser from the OOP mode sideband, when $\delta$ is zero, then the Raman transition is directly red and blue sideband transitions. For the M-S gate, $\delta$ should match to the sideband Rabi frequency and determines the duration of the M-S interaction as $1/\delta$. $\delta=22.0$ kHz here. (d), Frequencies of vibrational modes of a single \Yb\ and a single \Ba\ ions. Axial OOP mode is used for the M{\o}lmer-S{\o}rensen (M-S) interaction. IP stands for in-phase mode.}
	\label{figsys}
\end{figure}

The test of inequality (\ref{ineq}) is performed on a two-qubit system in which each qubit is encoded in a different atomic ion. One of \Yb\ and the other of \Ba\,, both trapped in a four-rod Paul trap \cite{WolfgangRMP}
, as shown in Fig.~\ref{figsys}. The first qubit is encoded in two hyperfine levels of the $^2S_{1/2}$ manifold of the \Yb\ ion. The corresponding states are denoted $\ket{0}_{\rm{Yb}}\equiv\ket{F=0,m_F=0}$ and $\ket{1}_{\rm{Yb}}\equiv\ket{F=1,m_F=0}$. The energy gap between the two states is $f_{\textrm{Yb}}=12.64281\Unit{GHz}$. The second qubit is encoded in the two Zeeman levels of the $^2S_{1/2}$ manifold of the \Ba\, ion. The corresponding states are denoted $\ket{0}_{\rm{Ba}}\equiv\ket{m=1/2}$ and $\ket{1}_{\rm{Ba}}\equiv\ket{m=-1/2}$. The energy gap is $f_{\textrm{Ba}}=16.8 \Unit{MHz}$ in an external magnetic field of $6.0\Unit{Gauss}$.

The two-ion system is initially prepared in state $\ket{\psi}=\frac{1}{\sqrt{2}}(\ket{00}+i\ket{11})$. The state of each qubit can be measured with a fluorescence detection technique. For the \Yb\, ion, the cyclic transition between $\ket{F=1}$ states in $^2S_{1/2}$ and $\ket{F=0,m_F=0}$ in $^2P_{1/2}$ is excited with a $370\Unit{nm}$ laser beam so that only $\ket{1}_{\rm{Yb}}$ scatters photons. The error of detecting $\ket{1}_{\rm{Yb}}$ for $\ket{0}_{\rm{Yb}}$ is 0.96\% and the other error is 2.25\%. For the \Ba\, ion, we first transfer the population of $\ket{0}_{\rm{Ba}}$ to $^2D_{5/2}$ with a $1762\Unit{nm}$ laser beam before exciting the $493\Unit{nm}$ transition between $^2S_{1/2}$ and $^2P_{1/2}$ levels. The error of detecting $\ket{1}_{\rm{Ba}}$ for $\ket{0}_{\rm{Ba}}$ is 2.10\% and the other error is below 0.01\%. A $1064\Unit{nm}$ picosecond pulsed laser is used for the coherent quantum operations of the two qubits. Two beams from its $532\Unit{nm}$ frequency-doubled output are used to generate a stimulated Raman process to control the \Ba\, ion, and another two beams from its $355\Unit{nm}$ frequency-tripled output are used for the \Yb\, ion \cite{Hayes2010}. The schematic diagram of the arrangement of both Raman laser beams is shown in Fig.~\ref{figsys}(c).

\begin{figure}[htp]
	\centering
	\hspace{-40pt}
	\includegraphics[width=0.4\textwidth]{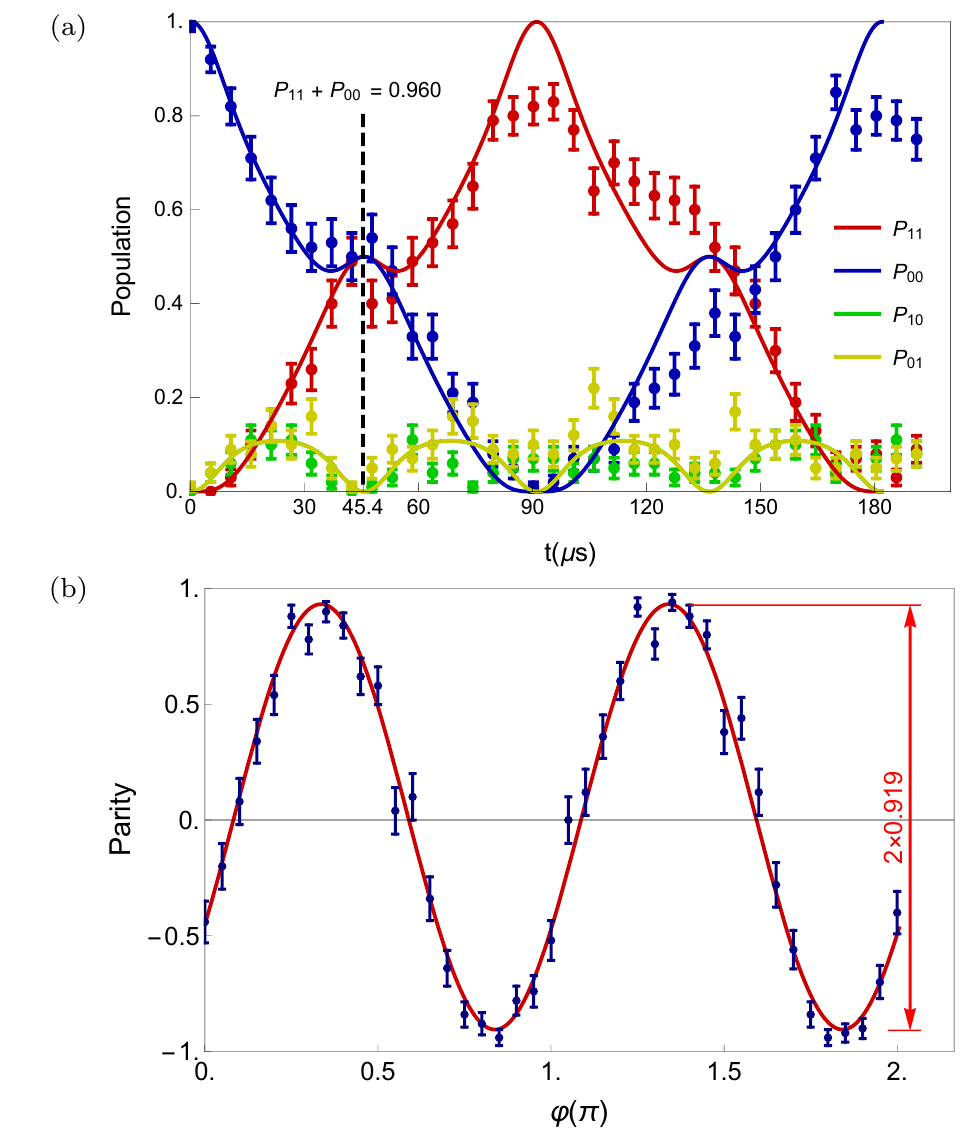}
	\caption{\textbf{Evolution of the M-S interaction and oscillation of parity signal.} Each data point is the average of 100 repetitions and all the error bars are standard deviations. (a) The time evolution of the M-S interaction. $P_{ij}$ is the population of state $\ket{ij}$, where $\ket{i,j}=\ket{i}_{\rm{Yb}}\ket{j}_{\rm{Ba}}$. The duration of a single M-S gate is 45.4 \textmu s and $P_{11}+P_{00} = 0.960 \pm 0.018$ at the end of the gate. (b) The parity scan of the entangled state. Parity of a state is defined as $P_{11}+P_{00}-P_{10}-P_{01}$, which is the population difference between the two qubits being in same or opposite states. Parity contrast is $0.919 \pm 0.021$.
		\label{figMSgate}}
\end{figure}

State $\ket{\psi}$ is generated through the M{\o}lmer-S{\o}rensen (M-S) interaction mediated by the axial out-of-phase (OOP) mode of the two ions with a frequency of $f_\textrm{z}=1.67\Unit{MHz}$ \cite{Inlek2017}.
The average phonon number of axial OOP mode is cooled down to ~0.04, and the IP mode is cooled down to ~0.11 with Doppler cooling, electromagnetically-induced-transparency (EIT) cooling \cite{Lechner2016}
and Raman sideband cooling \cite{Roos1999}.
We note that EIT cooling and Raman sideband cooling are performed only by \Ba\ ion, which sympathetically cools the \Yb\ ion. The time evolution of the M-S interaction is shown in Fig.~\ref{figMSgate}(a). After the M-S gate, we apply $\pi /2$ rotations to both ions with varying phases and obtain the parity oscillation signal as shown in Fig.~\ref{figMSgate}(b). According to the state population after the M-S gate and the contrast of the parity oscillation, we obtain a fidelity of the generated entangled state $\ket{\psi}$ of $0.939\pm0.014$.

Gate errors mainly come from parameter drifts due to the long time calibration process and imperfect cooling of axial IP mode.

\begin{table*}[htp]
	\begin{center}
	\newcommand{\sn}{\hphantom{-}}
	\caption{Experiment settings and results of mean values and standard error of the mean (SEM). Each setting repeats 10000 times. $\EOij{i}{j}$ is the expectation value of observable $\hat{O}_i$ measured jointly with observable $\hat{O}_j$.
	\label{tdata}}
		\begin{tabular}{c|c|c|c}
			$\{i,j\}$ & $\EOiOj{i}{j}$        & $\EOij{i}{j}$         & $\EOij{j}{i}$      \\
			\hline
			$\{0,1\}$ & $\sn 0.6164\pm0.0079$ & $\sn -0.0008\pm0.0100$ & $0.1096\pm0.0099$ \\
			$\{1,2\}$ & $\sn 0.625\pm0.0078$ & $0.1066\pm0.0099$    &  $0.1236\pm0.0099$\\
			$\{2,3\}$ & $\sn 0.6678\pm0.0074$ & $0.1356\pm0.0099$    & $\sn 0.1078\pm0.0099$ \\
			$\{3,0\}$ & $ -0.6166\pm0.0079$    & $0.1114\pm0.0099$    & $-0.0056\pm0.0100$
		\end{tabular}\\
		\end{center}
\end{table*}

{\em Results and analysis of the loopholes.---}After the generation of the entangled state, one of the four contexts is chosen and measured.
For each ion, a $\pi/2$ rotation is first performed to map the corresponding observable to the $\hat{\sigma}_z$ basis, and then the fluorescence detection is performed.
The experiment is repeated $40000$~times. The acquired data with standard errors are shown in Table~\ref{tdata}.

The validity of the assumption of outcome noncontextuality that leads to the bound of inequality (\ref{ineq}) relies on the assumption that measurements are sharp \cite{Spekkens2005}.
That is, they yield the same outcome when repeated and do not disturb measurements in the same context \cite{YuanXiao2014,Cabello2019Quantumcorrelations}.

In our experiment, measurement repeatability is checked 
by measuring the same observable two times in a single experimental run. For each observable, this is repeated $1000$~times.

\begin{figure}[htp]
	\centering
	\includegraphics[width=0.4\textwidth]{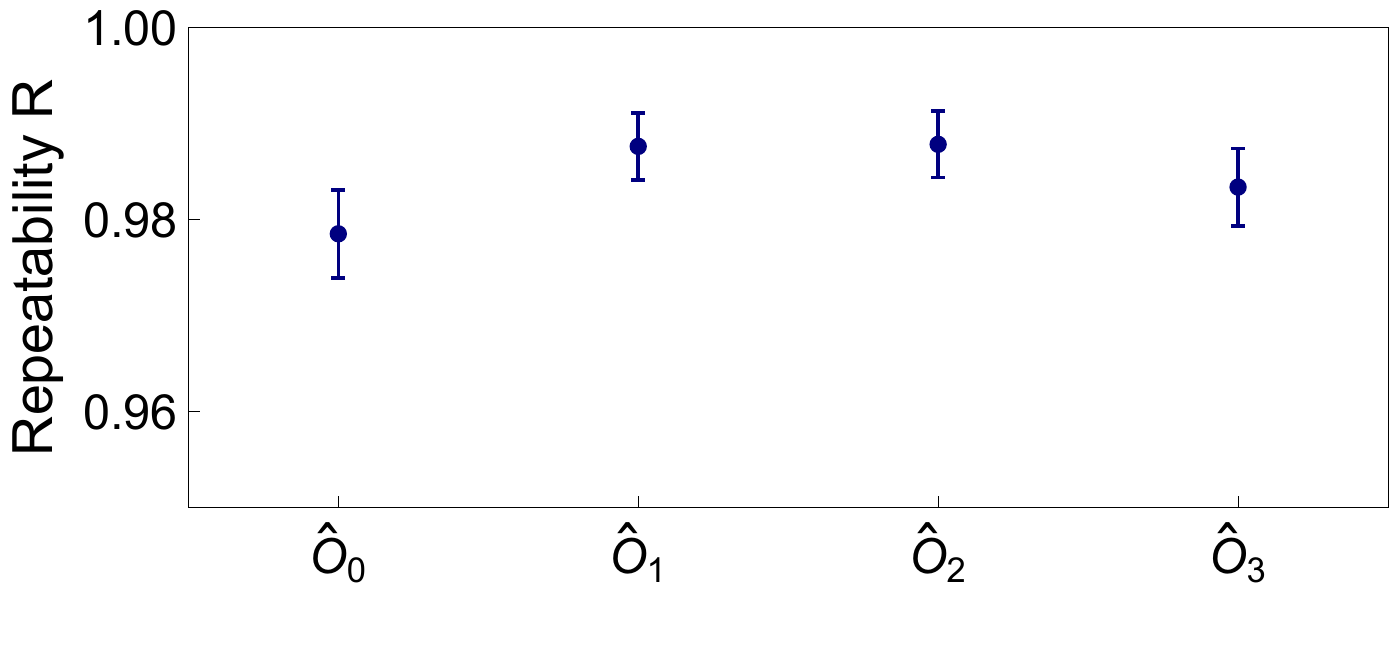}
	\caption{\textbf{Repeatability of the measurements.} The repeatability of each observable is tested 1000 times. Error bars are the standard error of the mean (SEM).
	}
	\label{figRepeatData}
\end{figure}

We define the repeatability $R_i$ as the fraction of measurement runs in which the observable $\hat{O}_i$ is measured twice and both of the outcomes are the same. For perfectly sharp measurements, $R_i$ should be $1$ for all $i=0\dots3$. In our experiment, the average value for the four observable is $98.4\%\pm 0.4\%$. Within our experimental error bars, the imperfection in the repeatability can be explained by mainly the detection infidelity of dark states for both ions, which is $1.5\%\pm0.4\%$ in average. The repeatability for each of the four observables is shown in Fig.~\ref{figRepeatData}. The sequence used for testing the repeatability is discussed in the Materials and Methods.

Nondisturbance between measurements in the same context is enforced by choosing trivially compatible observables. The deviation form perfect nondisturbance is attributable to finite statistics.

Repeatability could be further improved, up to $99.9\%$, by adopting closer to ideal equipment. To show that a repeatability $\sim 98.4\%$ is enough to close the sharpness loophole, we consider three types of noncontextual models that exploit this imperfection to increase the value of ${\cal C}$ beyond the limit for the ideal case. For these models, the bound of inequality~(\ref{ineq}) has to be corrected as follows:
\begin{equation}
\begin{split}
\mathcal{C}=&\EOiOj{0}{1}+\EOiOj{1}{2}+\EOiOj{2}{3}-\EOiOj{3}{0} \leq2 + \varepsilon,\\
\end{split}
\label{ineq2}
\end{equation}
where $\varepsilon$ depends on the way the noncontextual models may take advantage of the imperfections. We focus on three types of models.

The models considered in \cite{NDSC12}, based on the assumption that outcome noncontextuality holds only for a fraction $f$ of trials where the assumption of repeatability is satisfied for both measurements, while for the other fraction, $1-f$, the worst case scenario is assumed. That is, with probability $1-f$ the hidden variables can conspire to achieve the maximum algebraic value of ${\cal C}$. In our experiment, $f= 0.984^2 = 0.97$ and the maximum algebraic value of ${\cal C}$ is $4$. Therefore, $\varepsilon=0.06$.

The ``maximally noncontextual models'' \cite{Kujala2015,Malinowski&HomeJP&Cabello2018}
defined as those in which outcome noncontextuality holds with the maximum probability allowed by the observed marginals. That is, models that are only as conspiratorial as needed to account for the disturbance observed in the marginals. For these models, $\varepsilon=\sum_{i=0}^{3}\left|\EOij{i}{i\oplus 1}-\EOij{i}{i\ominus 1}\right|$, $\oplus$ is right shift $(0 \mapsto 1 \mapsto 2 \mapsto 3 \mapsto 0)$ and $\ominus$ is left shift $(0 \leftmapsto 1 \leftmapsto 2 \leftmapsto 3 \leftmapsto 0)$. $\EOiOj{i}{j}$ is the correlation between observable $\hat{O}_i$ and $\hat{O}_j$, and $\EOij{i}{j}$ is the expectation value of observable $\hat{O}_i$ measured jointly with observable $\hat{O}_j$. Using the results in Table~\ref{tdata}, for these models $\varepsilon=0.023 \pm 0.027$.

In addition, we consider the models \cite{Kirchmair&Cabello&Blatt2009,Guhne&Cabello2010} which apply to experiments with sequential {\em incompatible} measurements. In this case, the experimentally observed repeatability is used to estimate the disturbance that a measurement can cause to the result of the measurement performed afterwards and correct the bound for the ideal case. With our repeatability, these models lead to $\varepsilon=0.128$ (see \cite{Kirchmair&Cabello&Blatt2009,Guhne&Cabello2010} for details).

Using the data in Table~\ref{tdata} to evaluate $\mathcal{C}$ in inequality (\ref{ineq2}), we obtain $\mathcal{C}=2.526\pm 0.016$,
which corresponds to a violation of inequality~(\ref{ineq2}) for any of the models considered. Therefore, our experiment rules out noncontextual models maximally exploiting the lack of perfect repeatability, maximally noncontextual models, and we even consider a model which takes advantage of a lack of compatibility, which does not apply to our system.

To close the compatibility loophole, we map trivially compatible observables on separated ions of different species. Measurements on each ion use different operation laser wavelengths, fluorescence wavelengths, and detectors, as shown in Fig.~\ref{figsys}. The 355 nm laser beams perform coherent operations on the \Yb\ ion, while the 532 nm laser beams perform coherent operations on the \Ba\ ion. Although, in principle, the laser beams can also influence the ``wrong'' ion, this disturbance is too small to be detected, as it affects $\mathcal{C}$ at the level of $10^{-6}$ (see Materials and Methods).

To close the detection loophole, we adopt a scheme of 100$\%$ detection efficiency, which produces two measurement outcomes in every trial of the experiment. Therefore, the assumption of fair sampling \cite{Pearle1970} is not needed and the mere violation of the noncontextuality inequality (\ref{ineq2}) is enough to single out noncontextual models. However, due to the detection-infidelity, this strategy leads to a reduction of the violation of inequality (\ref{ineq2}) with respect to the one predicted by quantum mechanics for ideal equipment, $2 \sqrt{2}\sim 2.828$ \cite{cirel1980quantum}.

\section{Discussion}
Our experiment demonstrates that, as predicted by quantum mechanics, neither the persistency of a result when a measurement is repeated nor the observation that measurements in the same trial are not disturbing each other (as all of them yield the same outcome) imply that measurements reveal ``properties'' possessed by the systems. Our experiment shows, beyond any reasonable doubt, that nature allows for correlations between the outcomes of sharp measurements that cannot be explained by models based on the assumption of outcome noncontextuality.
This result is contrary to the deeply-rooted conception in science that persistency and repeatability of results imply the existence of properties revealed by the measurements. Our test is ``loophole-free'' in the sense that it closes simultaneously the main loopholes affecting previous contextuality tests. To the best of our knowledge, no other loopholes have been pointed out for KS contextuality experiments. However, in principle, there could be more loopholes. Inspiration for identifying them can be obtained in the following review paper on loopholes for Bell nonlocality experiments \cite{larsson2014loopholes}.

One could have argued that the only way to guarantee perfect compatibility is to spacelike separate the measurements. However, the same technical reasons (e.g., the finiteness of the experimental statistics and the impossibility of implementing the same measurement twice) that would make perfect non-disturbance and thus perfect compatibility impossible in a spatially (but not spacelike) separated experiment would also prevent any experiment with spacelike separation to achieve perfect compatibility. In this sense, an experiment with spatial (but not spacelike) separation in which the deviation from the non-disturbance condition is statistically negligible is as free of the compatibility loophole as any experiment can be. On the other hand, both in classical mechanics and in nonrelativistic quantum mechanics, two observables, $A$ on system 1 and $B$ on a spatially separated system 2, are trivially compatible as there is a third observable $C$ (which in this case is trivial as it can be measured by performing a measurement of $A$ on system 1 and a measurement of $B$ on system 2) and functions $f$ and $g$ such that $A=f(C)$ and $B=g(C)$ thus an outcome can be ascribed to both $A$ and $B$ by a single measurement of $C$. This argument has been used in previous proposals and experiments closing the compatibility loophole \cite{Cabello2011qutrit,SMC15,HuXiaomin2016,Borges&Cabello2014}.

Our results have direct implications to quantum algorithms and protocols running on devices where the assumption of locality cannot be made, as it is the case of quantum computers \cite{kleinmann2012optimal}. These devices are not large enough to allow for spacelike related events that justify the assumption of locality. There, the possibility of producing loophole-free contextual correlations for sharp measurements allows,  without relying on locality, for testing whether a claimed quantum computer is truly quantum \cite{barz2012demonstration}, characterizing quantum systems \cite{reichardt2013classical,Bharti2019}, self-testing quantum random number generation \cite{Lunghi2015}, and blind quantum computation \cite{barz2012demonstration}, among other applications.

\section{Materials and Methods}
\subsection*{Repeatability test}
\begin{figure}[htp]
	\centering
	\includegraphics[width=0.4\textwidth]{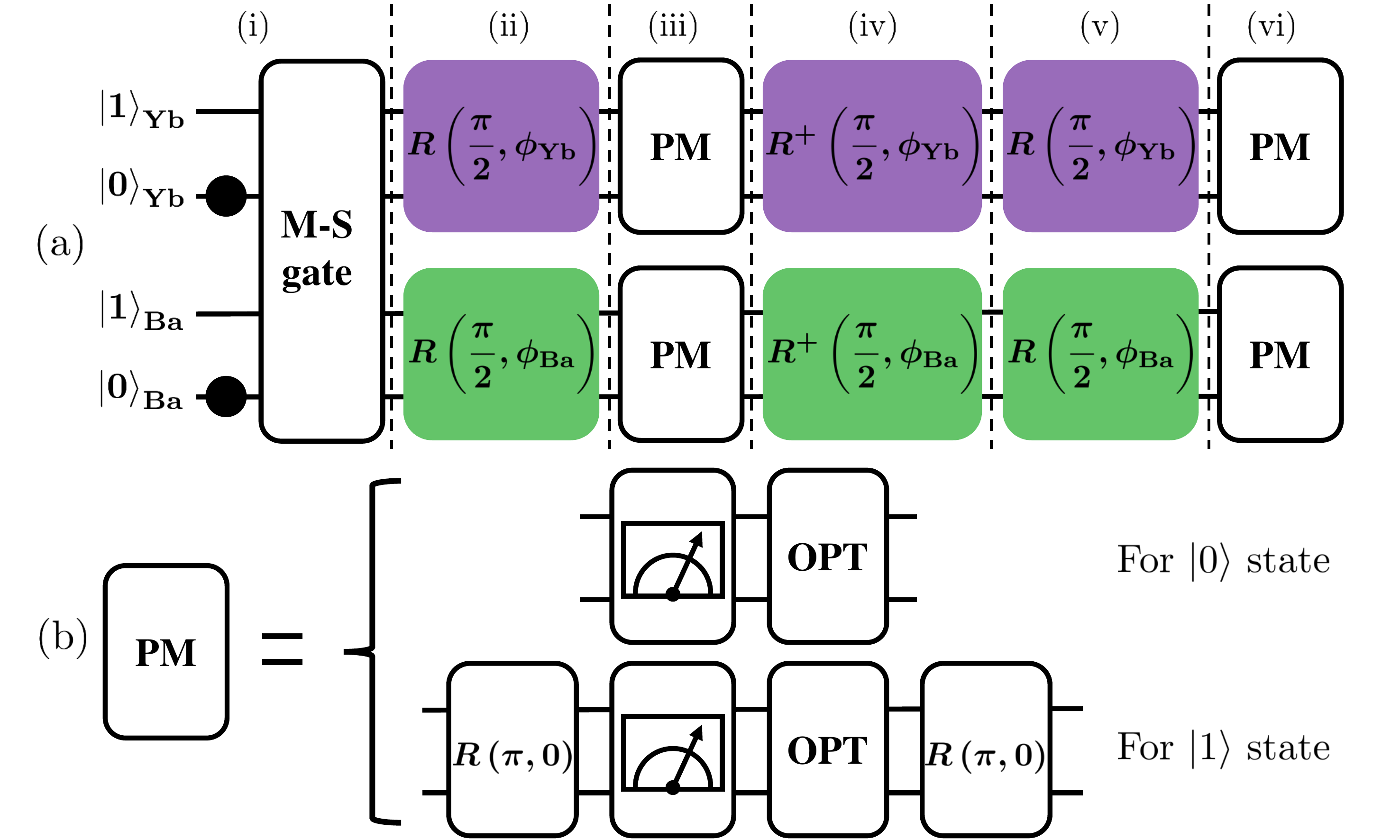}
	\caption{
	\textbf{Sequence used in the test of the repeatability of the measurements.} (a)The whole sequence for the test. The sequence includes six steps: (i) Pump two qubits to $\ket{0}$ and then prepare the entangled state with a M-S gate. (ii) Rotate the measurement basis $\hat{\sigma}_z$ to the observable basis $\hat{O}_i=R^{+}(\frac{\pi}{2},\phi)\hat{\sigma}_zR(\frac{\pi}{2},\phi)$, where
	$\phi_{\rm{Yb}}=\frac{5\pi}{4}$ and $\frac{3\pi}{4}$ for observables $\hat{O}_0$ and $\hat{O}_2$, $\phi_{\rm{Ba}}=\frac{3\pi}{2}$ and $\pi$ for observables $\hat{O}_1$ and $\hat{O}_3$. (iii) Projective measurement (PM). (iv) Rotate the measurement basis back. (v) Rotate the measurement basis to the observable basis again. (vi) PM again. $R\left(\frac{\pi}{2},\phi_{\textrm{Yb}}\right)$ in the purple box and $R\left(\frac{\pi}{2},\phi_\textrm{Ba}\right)$ in the green box are $\pi/2$ rotations of the \Yb and \Ba~ qubits, respectively. Only rotations in the purple box will be applied when observable $\hat{O}_0$ or $\hat{O}_2$ are measured since they only performed on the \Yb~ion. Similarly, only rotations in the green box will be applied for observables $\hat{O}_1$ and $\hat{O}_3$. (b)Scheme of projective measurement (PM) with post-selection for the test of repeatability. The PM measurement of the $\ket{0}$ state is realized by a single-shot fluorescence measurement and an optical pumping (OPT) pulse. The OPT pulse is used to recover the measured $\ket{0}$ state. The PM measurement of the $\ket{1}$ state is realized by $\pi$-rotation before and after the single-shot fluorescence measurement and optical pumping pulse. Here for the measurement of both $\ket{0}$ and $\ket{1}$, the results of no fluorescence are selected.}
	\label{figRepeatSeq}
\end{figure}

We perform the repeatability test by measuring the same observable two times in a single experimental run. The sequence used for testing the repeatability is discussed in Fig.~\ref{figRepeatSeq}. 
A single qubit rotation by $\theta$ about the $\cos(\phi)\hat{\sigma}_x + \sin(\phi)\hat{\sigma}_y$ axis is defined as
\begin{equation}
R(\theta,\phi)=
\left(
  \begin{array}{cc}
    \cos(\frac{\theta}{2}) & -ie^{-i\phi}\sin(\frac{\theta}{2}) \\
    -ie^{i\phi}\sin(\frac{\theta}{2}) & \cos(\frac{\theta}{2}) \\
  \end{array}
\right).
\end{equation}
Our projective measurements require different sequences depending on the detected state because the fluorescence detection for the $\ket{0}$ state (dark state) is ideal but it is not ideal for the $\ket{1}$ state (bright state). The bright-state detection is not ideal due to the leakage to other states outside qubit space. We address this problem by adopting post-selection technique, where only results of the dark state are collected. We first collect the data of only the $\ket{0}$ state (dark state) and abandon the data of the $\ket{1}$ (bright state). We test the case of $\ket{1}$ similar to that of $\ket{0}$ by inverting the $\ket{1}$ to the $\ket{0}$. We note that there is no fundamental problem of post-selection for the repeatability test, since we identify the not ideal data, whose first outcome is $\ket{1}$ (bright state), just by looking at the first outcome and without using any information about the second measurement setting or outcome and exclude them from the experiment.

\subsection*{Crosstalk between qubits}
\begin{figure}[htp]
    \centering
    \includegraphics[width=0.4\textwidth]{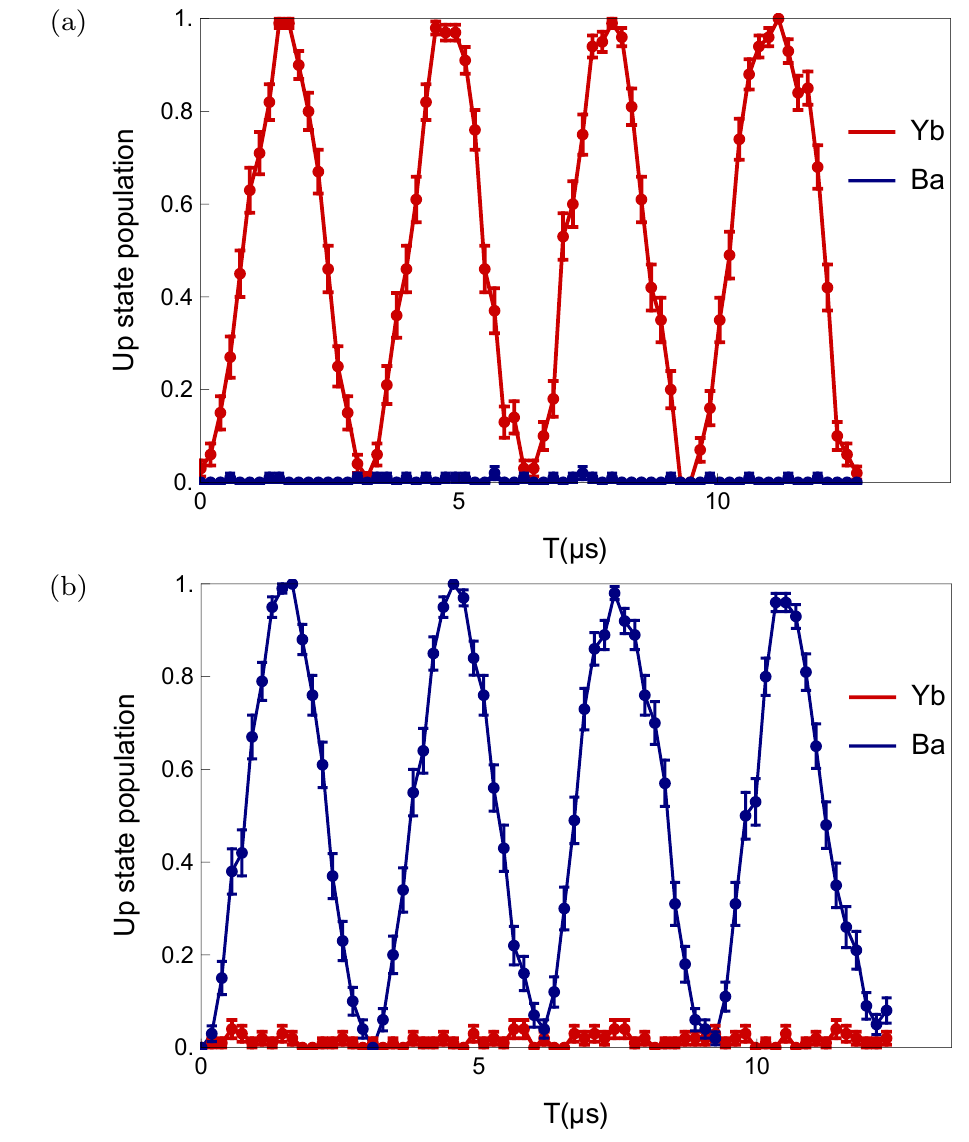}
    \caption{\textbf{Experimental test of the crosstalk between qubits.} Each point is repeated 100 times. Error bars are the standard error of the mean (SEM). (a) The up state population of both ions when 355 nm Raman laser is applied. (b) The up state population of both ions when 532 nm Raman laser is applied.}
    \label{figCrosstalk}
\end{figure}

The 355 nm and 532 nm Raman lasers are designed to drive the transition of the \Yb and \Ba~ions, respectively. However, in principle, they can also drive the \Ba~ and \Yb~ions, respectively. But this crosstalk is quite small. As shown in Fig~\ref{figCrosstalk}, When one of the Raman lasers is applied to the system, the ``wrong" ion doesn't have any excitation other than fluctuations caused by detection errors. 
This crosstalk is too small to be detected but easy to be estimated in theory. For that, we assume, firstly, that the pulse laser comb differences are resonant with the qubit transitions and only consider the energy structure of the ions and laser wavelength. The Raman transition strengths of \Yb~and \Ba~ on the laser wavelength and are \cite{Inlek2017}

\begin{align}
\Omega_{\rm{Yb}}&=\frac{I}{12}\left(-\frac{k_{1}}{\Delta_1}+\frac{k_{2}}{\Delta_2}\right),\\
\Omega_{\rm{Ba}} & =\frac{\sqrt{2}I}{12}\left(-\frac{k_1}{\Delta_1}+\frac{k_2}{\Delta_2}\right),
\end{align}
where subscripts 1 and 2 refer to the $\rm{P}_{1/2}$ and $\rm{P}_{3/2}$ levels, respectively. $I$ is the laser intensity and $k_i={\gamma_i^2}/{I_{sat,i}}$. $\gamma_i$, $I_{sat,i}$ and $\Delta_i$ are the natural linewidth, saturation intensity, and detuning for corresponding level. All related parameters  are shown in table \ref{Ionleveldata} \cite{Inlek2017}.
\begin{table*}[htp]
	\begin{center}
	\newcommand{\sn}{\hphantom{-}}
	\caption{Natural linewidth, saturation intensity, and laser detunings.
	\label{Ionleveldata}}
		\begin{tabular}{c|c|c|c|c}
			 Parameters &\multicolumn{2}{c|}{\Yb} &\multicolumn{2}{c}{\Ba}  \\
			\hline
			  Excited level &$^2P_{1/2}$ &$^2P_{3/2}$ &$^2P_{1/2}$ &$^2P_{3/2}$\\
			  Natural linewidth $\gamma/2\pi\mathrm{(MHz)}$ &19.7 &25.8 &15.1 &17.7 \\
			  Saturation intensity $I_{sat} \mathrm{(mW/cm^2)}$ &51.0 &95.1 &16.4 &35.7 \\
			  355 nm laser detunings $\Delta/2\pi\mathrm{(THz)}$ &34   &-66 &238 &187   \\
			  532 nm laser detunings $\Delta/2\pi\mathrm{(THz)}$  &-248 &-347  &-44 &-94 \\
			  $k~\mathrm{(MHz)^2*cm^2/mW}$  &7.61 &7.00 &13.90 &8.78 \\
		\end{tabular}
		\end{center}
\end{table*}

In our experiment, the transition strength $\Omega_{\rm{Yb,355}}=\Omega_{\rm{Ba,532}}=(2\pi)~0.18$ MHz, which leads to $I_{532}=6.86\times 10^{6}~\rm{mW/cm^2}$, $I_{355}=6.37\times 10^{6}~\rm{mW/cm^2}$. Then, the unwanted crosstalk transition strengths are
\begin{align}
|\Omega_{\rm{Yb},532}|&=\frac{I_{532}}{12}\left(\frac{k_1}{248~\rm{THz}}-\frac{k_2}{347~\rm{THz}}\right)\\
&=(2\pi)~0.006~\rm{MHz},\\
|\Omega_{\rm{Ba},355}| &=-\frac{\sqrt{2}I_{355}}{12}\left(-\frac{k_1}{187~\rm{THz}}+\frac{k_2}{238~\rm{THz}}\right)\\
&=(2\pi)~0.009~\rm{MHz}.
\end{align}
The two ions are assumed to be uniformly illuminated by the lasers. However, in the real experiment, both beams are aligned to their respective target ions, which further reduces the crosstalk.

Secondly, consider the comb difference between two lasers. The repetition rate of our pulse laser is 80.097 MHz. The frequency shift between two 355 (532) nm beam combs is 12.5 (16.3) MHz to meet the \Yb~(\Ba) qubit splitting of 12642.8 (16.3) MHz. Then, the undesirable Raman transitions have detuning of at least $|\Delta| /2\pi=16.8~\rm{MHz}-12.5~\rm{MHz}=4.3$ MHz for both \Ba~and \Yb~qubit transitions. These far-detuned couplings cause limited maximum population transfers in a single pulse of
\begin{align}
P_{\rm{max},\rm{Yb},532} & =\frac{\Omega_{\rm{Yb},532}^2}{\Delta^2+\Omega_{\rm{Yb},532}^2}=\frac{0.006^2}{4.3^2+0.006^2}=1.9\times 10^{-6}, \\
P_{\rm{max},\rm{Ba},355} & =\frac{\Omega_{\rm{Ba},355}^2}{\Delta^2+\Omega_{\rm{Ba},355}^2}=\frac{0.009^2}{4.3^2+0.009^2}=4.3\times 10^{-6}.
\end{align}
This amount of crosstalk is negligible for our experiment.

\section*{Acknowledgments}
We thank Costantino Budroni for comments.
\subsection*{Fundings}
This work was supported by the National Key Research and Development Program of China under Grants No.\ 2016YFA0301900 and No.\ 2016YFA0301901, the National Natural Science Foundation of China Grants No.\ 92065205, and No.\ 11974200, Project Qdisc (Project No.\ US-15097), with FEDER funds, and QuantERA grant SECRET, by MINECO (Project No.\ PCI2019-111885-2), and Guangdong Basic and Applied Basic Research Foundation Grant No. 2019A1515111135, and the Key-Area Research and Development Program of Guangdong Province Grant No. 2019B030330001.

\bibliography{Contextuality}

\begin{thebibliography}{65}%
\makeatletter
\providecommand \@ifxundefined [1]{%
 \@ifx{#1\undefined}
}%
\providecommand \@ifnum [1]{%
 \ifnum #1\expandafter \@firstoftwo
 \else \expandafter \@secondoftwo
 \fi
}%
\providecommand \@ifx [1]{%
 \ifx #1\expandafter \@firstoftwo
 \else \expandafter \@secondoftwo
 \fi
}%
\providecommand \natexlab [1]{#1}%
\providecommand \enquote  [1]{``#1''}%
\providecommand \bibnamefont  [1]{#1}%
\providecommand \bibfnamefont [1]{#1}%
\providecommand \citenamefont [1]{#1}%
\providecommand \href@noop [0]{\@secondoftwo}%
\providecommand \href [0]{\begingroup \@sanitize@url \@href}%
\providecommand \@href[1]{\@@startlink{#1}\@@href}%
\providecommand \@@href[1]{\endgroup#1\@@endlink}%
\providecommand \@sanitize@url [0]{\catcode `\\12\catcode `\$12\catcode
  `\&12\catcode `\#12\catcode `\^12\catcode `\_12\catcode `\%12\relax}%
\providecommand \@@startlink[1]{}%
\providecommand \@@endlink[0]{}%
\providecommand \url  [0]{\begingroup\@sanitize@url \@url }%
\providecommand \@url [1]{\endgroup\@href {#1}{\urlprefix }}%
\providecommand \urlprefix  [0]{URL }%
\providecommand \Eprint [0]{\href }%
\providecommand \doibase [0]{https://doi.org/}%
\providecommand \selectlanguage [0]{\@gobble}%
\providecommand \bibinfo  [0]{\@secondoftwo}%
\providecommand \bibfield  [0]{\@secondoftwo}%
\providecommand \translation [1]{[#1]}%
\providecommand \BibitemOpen [0]{}%
\providecommand \bibitemStop [0]{}%
\providecommand \bibitemNoStop [0]{.\EOS\space}%
\providecommand \EOS [0]{\spacefactor3000\relax}%
\providecommand \BibitemShut  [1]{\csname bibitem#1\endcsname}%
\let\auto@bib@innerbib\@empty
\bibitem [{\citenamefont {Klyachko}\ \emph {et~al.}(2008)\citenamefont
  {Klyachko}, \citenamefont {Can}, \citenamefont {Binicioglu},\ and\
  \citenamefont {Shumovsky}}]{Klyachko2008}%
  \BibitemOpen
  \bibfield  {author} {\bibinfo {author} {\bibfnamefont {A.~A.}\ \bibnamefont
  {Klyachko}}, \bibinfo {author} {\bibfnamefont {M.~A.}\ \bibnamefont {Can}},
  \bibinfo {author} {\bibfnamefont {S.}~\bibnamefont {Binicioglu}},\ and\
  \bibinfo {author} {\bibfnamefont {A.~S.}\ \bibnamefont {Shumovsky}},\
  }\bibfield  {title} {\bibinfo {title} {Simple test for hidden variables in
  spin-1 systems},\ }\href {https://doi.org/10.1103/PhysRevLett.101.020403}
  {\bibfield  {journal} {\bibinfo  {journal} {Phys. Rev. Lett.}\ }\textbf
  {\bibinfo {volume} {101}},\ \bibinfo {pages} {020403} (\bibinfo {year}
  {2008})}\BibitemShut {NoStop}%
\bibitem [{\citenamefont {Cabello}(2008)}]{Cabello2008exp}%
  \BibitemOpen
  \bibfield  {author} {\bibinfo {author} {\bibfnamefont {A.}~\bibnamefont
  {Cabello}},\ }\bibfield  {title} {\bibinfo {title} {Experimentally testable
  state-independent quantum contextuality},\ }\href
  {https://doi.org/10.1103/PhysRevLett.101.210401} {\bibfield  {journal}
  {\bibinfo  {journal} {Phys. Rev. Lett.}\ }\textbf {\bibinfo {volume} {101}},\
  \bibinfo {pages} {210401} (\bibinfo {year} {2008})}\BibitemShut {NoStop}%
\bibitem [{\citenamefont {Bell}(1966)}]{Bell1966}%
  \BibitemOpen
  \bibfield  {author} {\bibinfo {author} {\bibfnamefont {J.~S.}\ \bibnamefont
  {Bell}},\ }\bibfield  {title} {\bibinfo {title} {On the problem of hidden
  variables in quantum mechanics},\ }\href
  {https://doi.org/10.1103/revmodphys.38.447} {\bibfield  {journal} {\bibinfo
  {journal} {Rev. Mod. Phys.}\ }\textbf {\bibinfo {volume} {38}},\ \bibinfo
  {pages} {447} (\bibinfo {year} {1966})}\BibitemShut {NoStop}%
\bibitem [{\citenamefont {Kochen}\ and\ \citenamefont
  {Specker}(1967)}]{Kochen1967}%
  \BibitemOpen
  \bibfield  {author} {\bibinfo {author} {\bibfnamefont {S.}~\bibnamefont
  {Kochen}}\ and\ \bibinfo {author} {\bibfnamefont {E.~P.}\ \bibnamefont
  {Specker}},\ }\bibfield  {title} {\bibinfo {title} {The problem of hidden
  variables in quantum mechanics},\ }\href@noop {} {\bibfield  {journal}
  {\bibinfo  {journal} {J. Math. Mech.}\ }\textbf {\bibinfo {volume} {17}},\
  \bibinfo {pages} {59} (\bibinfo {year} {1967})}\BibitemShut {NoStop}%
\bibitem [{\citenamefont {Anders}\ and\ \citenamefont {Browne}(2009)}]{AB09}%
  \BibitemOpen
  \bibfield  {author} {\bibinfo {author} {\bibfnamefont {J.}~\bibnamefont
  {Anders}}\ and\ \bibinfo {author} {\bibfnamefont {D.~E.}\ \bibnamefont
  {Browne}},\ }\bibfield  {title} {\bibinfo {title} {Computational power of
  correlations},\ }\href {https://doi.org/10.1103/PhysRevLett.102.050502}
  {\bibfield  {journal} {\bibinfo  {journal} {Phys. Rev. Lett.}\ }\textbf
  {\bibinfo {volume} {102}},\ \bibinfo {pages} {050502} (\bibinfo {year}
  {2009})}\BibitemShut {NoStop}%
\bibitem [{\citenamefont {Raussendorf}(2013)}]{Raussendorf13}%
  \BibitemOpen
  \bibfield  {author} {\bibinfo {author} {\bibfnamefont {R.}~\bibnamefont
  {Raussendorf}},\ }\bibfield  {title} {\bibinfo {title} {Contextuality in
  measurement-based quantum computation},\ }\href
  {https://doi.org/10.1103/PhysRevA.88.022322} {\bibfield  {journal} {\bibinfo
  {journal} {Phys. Rev. A}\ }\textbf {\bibinfo {volume} {88}},\ \bibinfo
  {pages} {022322} (\bibinfo {year} {2013})}\BibitemShut {NoStop}%
\bibitem [{\citenamefont {Howard}\ \emph {et~al.}(2014)\citenamefont {Howard},
  \citenamefont {Wallman}, \citenamefont {Veitch},\ and\ \citenamefont
  {Emerson}}]{Howard2014}%
  \BibitemOpen
  \bibfield  {author} {\bibinfo {author} {\bibfnamefont {M.}~\bibnamefont
  {Howard}}, \bibinfo {author} {\bibfnamefont {J.}~\bibnamefont {Wallman}},
  \bibinfo {author} {\bibfnamefont {V.}~\bibnamefont {Veitch}},\ and\ \bibinfo
  {author} {\bibfnamefont {J.}~\bibnamefont {Emerson}},\ }\bibfield  {title}
  {\bibinfo {title} {Contextuality supplies the ``magic'' for quantum
  computation},\ }\href {https://doi.org/10.1038/nature13460} {\bibfield
  {journal} {\bibinfo  {journal} {Nature}\ }\textbf {\bibinfo {volume} {510}},\
  \bibinfo {pages} {351} (\bibinfo {year} {2014})}\BibitemShut {NoStop}%
\bibitem [{\citenamefont {Delfosse}\ \emph {et~al.}(2015)\citenamefont
  {Delfosse}, \citenamefont {Allard~Guerin}, \citenamefont {Bian},\ and\
  \citenamefont {Raussendorf}}]{DGBR15}%
  \BibitemOpen
  \bibfield  {author} {\bibinfo {author} {\bibfnamefont {N.}~\bibnamefont
  {Delfosse}}, \bibinfo {author} {\bibfnamefont {P.}~\bibnamefont
  {Allard~Guerin}}, \bibinfo {author} {\bibfnamefont {J.}~\bibnamefont
  {Bian}},\ and\ \bibinfo {author} {\bibfnamefont {R.}~\bibnamefont
  {Raussendorf}},\ }\bibfield  {title} {\bibinfo {title} {Wigner function
  negativity and contextuality in quantum computation on rebits},\ }\href
  {https://doi.org/10.1103/PhysRevX.5.021003} {\bibfield  {journal} {\bibinfo
  {journal} {Phys. Rev. X}\ }\textbf {\bibinfo {volume} {5}},\ \bibinfo {pages}
  {021003} (\bibinfo {year} {2015})}\BibitemShut {NoStop}%
\bibitem [{\citenamefont {Bravyi}\ \emph {et~al.}(2018)\citenamefont {Bravyi},
  \citenamefont {Gosset},\ and\ \citenamefont {K\"onig}}]{Bravyi18}%
  \BibitemOpen
  \bibfield  {author} {\bibinfo {author} {\bibfnamefont {S.}~\bibnamefont
  {Bravyi}}, \bibinfo {author} {\bibfnamefont {D.}~\bibnamefont {Gosset}},\
  and\ \bibinfo {author} {\bibfnamefont {R.}~\bibnamefont {K\"onig}},\
  }\bibfield  {title} {\bibinfo {title} {Quantum advantage with shallow
  circuits},\ }\href {https://doi.org/10.1126/science.aar3106} {\bibfield
  {journal} {\bibinfo  {journal} {Science}\ }\textbf {\bibinfo {volume}
  {362}},\ \bibinfo {pages} {308} (\bibinfo {year} {2018})}\BibitemShut
  {NoStop}%
\bibitem [{\citenamefont {Michler}\ \emph {et~al.}(2000)\citenamefont
  {Michler}, \citenamefont {Weinfurter},\ and\ \citenamefont
  {\ifmmode~\dot{Z}\else \.{Z}\fi{}ukowski}}]{MWZ00}%
  \BibitemOpen
  \bibfield  {author} {\bibinfo {author} {\bibfnamefont {M.}~\bibnamefont
  {Michler}}, \bibinfo {author} {\bibfnamefont {H.}~\bibnamefont
  {Weinfurter}},\ and\ \bibinfo {author} {\bibfnamefont {M.}~\bibnamefont
  {\ifmmode~\dot{Z}\else \.{Z}\fi{}ukowski}},\ }\bibfield  {title} {\bibinfo
  {title} {Experiments towards falsification of noncontextual hidden variable
  theories},\ }\href {https://doi.org/10.1103/PhysRevLett.84.5457} {\bibfield
  {journal} {\bibinfo  {journal} {Phys. Rev. Lett.}\ }\textbf {\bibinfo
  {volume} {84}},\ \bibinfo {pages} {5457} (\bibinfo {year}
  {2000})}\BibitemShut {NoStop}%
\bibitem [{\citenamefont {Huang}\ \emph {et~al.}(2003)\citenamefont {Huang},
  \citenamefont {Li}, \citenamefont {Zhang}, \citenamefont {Pan},\ and\
  \citenamefont {Guo}}]{HLZPG03}%
  \BibitemOpen
  \bibfield  {author} {\bibinfo {author} {\bibfnamefont {Y.-F.}\ \bibnamefont
  {Huang}}, \bibinfo {author} {\bibfnamefont {C.-F.}\ \bibnamefont {Li}},
  \bibinfo {author} {\bibfnamefont {Y.-S.}\ \bibnamefont {Zhang}}, \bibinfo
  {author} {\bibfnamefont {J.-W.}\ \bibnamefont {Pan}},\ and\ \bibinfo {author}
  {\bibfnamefont {G.-C.}\ \bibnamefont {Guo}},\ }\bibfield  {title} {\bibinfo
  {title} {Experimental test of the kochen-specker theorem with single
  photons},\ }\href {https://doi.org/10.1103/PhysRevLett.90.250401} {\bibfield
  {journal} {\bibinfo  {journal} {Phys. Rev. Lett.}\ }\textbf {\bibinfo
  {volume} {90}},\ \bibinfo {pages} {250401} (\bibinfo {year}
  {2003})}\BibitemShut {NoStop}%
\bibitem [{\citenamefont {Amselem}\ \emph {et~al.}(2009)\citenamefont
  {Amselem}, \citenamefont {R{\r{a}}dmark}, \citenamefont {Bourennane},\ and\
  \citenamefont {Cabello}}]{Amselem&Cabello2009}%
  \BibitemOpen
  \bibfield  {author} {\bibinfo {author} {\bibfnamefont {E.}~\bibnamefont
  {Amselem}}, \bibinfo {author} {\bibfnamefont {M.}~\bibnamefont
  {R{\r{a}}dmark}}, \bibinfo {author} {\bibfnamefont {M.}~\bibnamefont
  {Bourennane}},\ and\ \bibinfo {author} {\bibfnamefont {A.}~\bibnamefont
  {Cabello}},\ }\bibfield  {title} {\bibinfo {title} {State-independent quantum
  contextuality with single photons},\ }\href
  {https://doi.org/10.1103/PhysRevLett.103.160405} {\bibfield  {journal}
  {\bibinfo  {journal} {Phys. Rev. Lett.}\ }\textbf {\bibinfo {volume} {103}},\
  \bibinfo {pages} {160405} (\bibinfo {year} {2009})}\BibitemShut {NoStop}%
\bibitem [{\citenamefont {Lapkiewicz}\ \emph {et~al.}(2011)\citenamefont
  {Lapkiewicz}, \citenamefont {Li}, \citenamefont {Schaeff}, \citenamefont
  {Langford}, \citenamefont {Ramelow}, \citenamefont {Wie\'sniak},\ and\
  \citenamefont {Zeilinger}}]{Lapkiewicz2011}%
  \BibitemOpen
  \bibfield  {author} {\bibinfo {author} {\bibfnamefont {R.}~\bibnamefont
  {Lapkiewicz}}, \bibinfo {author} {\bibfnamefont {P.}~\bibnamefont {Li}},
  \bibinfo {author} {\bibfnamefont {C.}~\bibnamefont {Schaeff}}, \bibinfo
  {author} {\bibfnamefont {N.~K.}\ \bibnamefont {Langford}}, \bibinfo {author}
  {\bibfnamefont {S.}~\bibnamefont {Ramelow}}, \bibinfo {author} {\bibfnamefont
  {M.}~\bibnamefont {Wie\'sniak}},\ and\ \bibinfo {author} {\bibfnamefont
  {A.}~\bibnamefont {Zeilinger}},\ }\bibfield  {title} {\bibinfo {title}
  {Experimental non-classicality of an indivisible quantum system},\ }\href
  {https://doi.org/10.1038/nature10119} {\bibfield  {journal} {\bibinfo
  {journal} {Nature}\ }\textbf {\bibinfo {volume} {474}},\ \bibinfo {pages}
  {490} (\bibinfo {year} {2011})}\BibitemShut {NoStop}%
\bibitem [{\citenamefont {Zu}\ \emph {et~al.}(2012)\citenamefont {Zu},
  \citenamefont {Wang}, \citenamefont {Deng}, \citenamefont {Chang},
  \citenamefont {Liu}, \citenamefont {Hou}, \citenamefont {Yang},\ and\
  \citenamefont {Duan}}]{ZWDCLHYD12}%
  \BibitemOpen
  \bibfield  {author} {\bibinfo {author} {\bibfnamefont {C.}~\bibnamefont
  {Zu}}, \bibinfo {author} {\bibfnamefont {Y.-X.}\ \bibnamefont {Wang}},
  \bibinfo {author} {\bibfnamefont {D.-L.}\ \bibnamefont {Deng}}, \bibinfo
  {author} {\bibfnamefont {X.-Y.}\ \bibnamefont {Chang}}, \bibinfo {author}
  {\bibfnamefont {K.}~\bibnamefont {Liu}}, \bibinfo {author} {\bibfnamefont
  {P.-Y.}\ \bibnamefont {Hou}}, \bibinfo {author} {\bibfnamefont {H.-X.}\
  \bibnamefont {Yang}},\ and\ \bibinfo {author} {\bibfnamefont {L.-M.}\
  \bibnamefont {Duan}},\ }\bibfield  {title} {\bibinfo {title}
  {State-independent experimental test of quantum contextuality in an
  indivisible system},\ }\href {https://doi.org/10.1103/PhysRevLett.109.150401}
  {\bibfield  {journal} {\bibinfo  {journal} {Phys. Rev. Lett.}\ }\textbf
  {\bibinfo {volume} {109}},\ \bibinfo {pages} {150401} (\bibinfo {year}
  {2012})}\BibitemShut {NoStop}%
\bibitem [{\citenamefont {D'Ambrosio}\ \emph {et~al.}(2013)\citenamefont
  {D'Ambrosio}, \citenamefont {Herbauts}, \citenamefont {Amselem},
  \citenamefont {Nagali}, \citenamefont {Bourennane}, \citenamefont
  {Sciarrino},\ and\ \citenamefont {Cabello}}]{DHANBSC13}%
  \BibitemOpen
  \bibfield  {author} {\bibinfo {author} {\bibfnamefont {V.}~\bibnamefont
  {D'Ambrosio}}, \bibinfo {author} {\bibfnamefont {I.}~\bibnamefont
  {Herbauts}}, \bibinfo {author} {\bibfnamefont {E.}~\bibnamefont {Amselem}},
  \bibinfo {author} {\bibfnamefont {E.}~\bibnamefont {Nagali}}, \bibinfo
  {author} {\bibfnamefont {M.}~\bibnamefont {Bourennane}}, \bibinfo {author}
  {\bibfnamefont {F.}~\bibnamefont {Sciarrino}},\ and\ \bibinfo {author}
  {\bibfnamefont {A.}~\bibnamefont {Cabello}},\ }\bibfield  {title} {\bibinfo
  {title} {Experimental implementation of a kochen-specker set of quantum
  tests},\ }\href {https://doi.org/10.1103/PhysRevX.3.011012} {\bibfield
  {journal} {\bibinfo  {journal} {Phys. Rev. X}\ }\textbf {\bibinfo {volume}
  {3}},\ \bibinfo {pages} {011012} (\bibinfo {year} {2013})}\BibitemShut
  {NoStop}%
\bibitem [{\citenamefont {Ahrens}\ \emph {et~al.}(2013)\citenamefont {Ahrens},
  \citenamefont {Amselem}, \citenamefont {Cabello},\ and\ \citenamefont
  {Bourennane}}]{AACB13}%
  \BibitemOpen
  \bibfield  {author} {\bibinfo {author} {\bibfnamefont {J.}~\bibnamefont
  {Ahrens}}, \bibinfo {author} {\bibfnamefont {E.}~\bibnamefont {Amselem}},
  \bibinfo {author} {\bibfnamefont {A.}~\bibnamefont {Cabello}},\ and\ \bibinfo
  {author} {\bibfnamefont {M.}~\bibnamefont {Bourennane}},\ }\bibfield  {title}
  {\bibinfo {title} {Two fundamental experimental tests of nonclassicality with
  qutrits},\ }\href {https://doi.org/10.1038/srep02170} {\bibfield  {journal}
  {\bibinfo  {journal} {Sci. Rep.}\ }\textbf {\bibinfo {volume} {3}},\ \bibinfo
  {pages} {1} (\bibinfo {year} {2013})}\BibitemShut {NoStop}%
\bibitem [{\citenamefont {Marques}\ \emph {et~al.}(2014)\citenamefont
  {Marques}, \citenamefont {Ahrens}, \citenamefont {Nawareg}, \citenamefont
  {Cabello},\ and\ \citenamefont {Bourennane}}]{MANCB14}%
  \BibitemOpen
  \bibfield  {author} {\bibinfo {author} {\bibfnamefont {B.}~\bibnamefont
  {Marques}}, \bibinfo {author} {\bibfnamefont {J.}~\bibnamefont {Ahrens}},
  \bibinfo {author} {\bibfnamefont {M.}~\bibnamefont {Nawareg}}, \bibinfo
  {author} {\bibfnamefont {A.}~\bibnamefont {Cabello}},\ and\ \bibinfo {author}
  {\bibfnamefont {M.}~\bibnamefont {Bourennane}},\ }\bibfield  {title}
  {\bibinfo {title} {Experimental observation of hardy-like quantum
  contextuality},\ }\href {https://doi.org/10.1103/PhysRevLett.113.250403}
  {\bibfield  {journal} {\bibinfo  {journal} {Phys. Rev. Lett.}\ }\textbf
  {\bibinfo {volume} {113}},\ \bibinfo {pages} {250403} (\bibinfo {year}
  {2014})}\BibitemShut {NoStop}%
\bibitem [{\citenamefont {Hasegawa}\ \emph {et~al.}(2003)\citenamefont
  {Hasegawa}, \citenamefont {Loidl}, \citenamefont {Badurek}, \citenamefont
  {Baron},\ and\ \citenamefont {Rauch}}]{YLB03}%
  \BibitemOpen
  \bibfield  {author} {\bibinfo {author} {\bibfnamefont {Y.}~\bibnamefont
  {Hasegawa}}, \bibinfo {author} {\bibfnamefont {R.}~\bibnamefont {Loidl}},
  \bibinfo {author} {\bibfnamefont {G.}~\bibnamefont {Badurek}}, \bibinfo
  {author} {\bibfnamefont {M.}~\bibnamefont {Baron}},\ and\ \bibinfo {author}
  {\bibfnamefont {H.}~\bibnamefont {Rauch}},\ }\bibfield  {title} {\bibinfo
  {title} {Violation of a bell-like inequality in single-neutron
  interferometry},\ }\href {https://doi.org/10.1038/nature0188} {\bibfield
  {journal} {\bibinfo  {journal} {Nature}\ }\textbf {\bibinfo {volume} {425}},\
  \bibinfo {pages} {45} (\bibinfo {year} {2003})}\BibitemShut {NoStop}%
\bibitem [{\citenamefont {Kirchmair}\ \emph {et~al.}(2009)\citenamefont
  {Kirchmair}, \citenamefont {Z\"ahringer}, \citenamefont {Gerritsma},
  \citenamefont {Kleinmann}, \citenamefont {G\"uhne}, \citenamefont {Cabello},
  \citenamefont {Blatt},\ and\ \citenamefont
  {Roos}}]{Kirchmair&Cabello&Blatt2009}%
  \BibitemOpen
  \bibfield  {author} {\bibinfo {author} {\bibfnamefont {G.}~\bibnamefont
  {Kirchmair}}, \bibinfo {author} {\bibfnamefont {F.}~\bibnamefont
  {Z\"ahringer}}, \bibinfo {author} {\bibfnamefont {R.}~\bibnamefont
  {Gerritsma}}, \bibinfo {author} {\bibfnamefont {M.}~\bibnamefont
  {Kleinmann}}, \bibinfo {author} {\bibfnamefont {O.}~\bibnamefont {G\"uhne}},
  \bibinfo {author} {\bibfnamefont {A.}~\bibnamefont {Cabello}}, \bibinfo
  {author} {\bibfnamefont {R.}~\bibnamefont {Blatt}},\ and\ \bibinfo {author}
  {\bibfnamefont {C.~F.}\ \bibnamefont {Roos}},\ }\bibfield  {title} {\bibinfo
  {title} {State-independent experimental test of quantum contextuality},\
  }\href {https://doi.org/10.1038/nature08172} {\bibfield  {journal} {\bibinfo
  {journal} {Nature}\ }\textbf {\bibinfo {volume} {460}},\ \bibinfo {pages}
  {494} (\bibinfo {year} {2009})}\BibitemShut {NoStop}%
\bibitem [{\citenamefont {Zhang}\ \emph {et~al.}(2013)\citenamefont {Zhang},
  \citenamefont {Um}, \citenamefont {Zhang}, \citenamefont {An}, \citenamefont
  {Wang}, \citenamefont {Deng}, \citenamefont {Shen}, \citenamefont {Duan},\
  and\ \citenamefont {Kim}}]{ZhangXiang2013}%
  \BibitemOpen
  \bibfield  {author} {\bibinfo {author} {\bibfnamefont {X.}~\bibnamefont
  {Zhang}}, \bibinfo {author} {\bibfnamefont {M.}~\bibnamefont {Um}}, \bibinfo
  {author} {\bibfnamefont {J.}~\bibnamefont {Zhang}}, \bibinfo {author}
  {\bibfnamefont {S.}~\bibnamefont {An}}, \bibinfo {author} {\bibfnamefont
  {Y.}~\bibnamefont {Wang}}, \bibinfo {author} {\bibfnamefont {D.~L.}\
  \bibnamefont {Deng}}, \bibinfo {author} {\bibfnamefont {C.}~\bibnamefont
  {Shen}}, \bibinfo {author} {\bibfnamefont {L.~M.}\ \bibnamefont {Duan}},\
  and\ \bibinfo {author} {\bibfnamefont {K.}~\bibnamefont {Kim}},\ }\bibfield
  {title} {\bibinfo {title} {State-independent experimental test of quantum
  contextuality with a single trapped ion},\ }\href
  {https://doi.org/10.1103/PhysRevLett.110.070401} {\bibfield  {journal}
  {\bibinfo  {journal} {Phys. Rev. Lett.}\ }\textbf {\bibinfo {volume} {110}},\
  \bibinfo {pages} {070401} (\bibinfo {year} {2013})}\BibitemShut {NoStop}%
\bibitem [{\citenamefont {Leupold}\ \emph {et~al.}(2018)\citenamefont
  {Leupold}, \citenamefont {Malinowski}, \citenamefont {Zhang}, \citenamefont
  {Negnevitsky}, \citenamefont {Cabello}, \citenamefont {Alonso},\ and\
  \citenamefont {Home}}]{CabelloJPhome2018}%
  \BibitemOpen
  \bibfield  {author} {\bibinfo {author} {\bibfnamefont {F.~M.}\ \bibnamefont
  {Leupold}}, \bibinfo {author} {\bibfnamefont {M.}~\bibnamefont {Malinowski}},
  \bibinfo {author} {\bibfnamefont {C.}~\bibnamefont {Zhang}}, \bibinfo
  {author} {\bibfnamefont {V.}~\bibnamefont {Negnevitsky}}, \bibinfo {author}
  {\bibfnamefont {A.}~\bibnamefont {Cabello}}, \bibinfo {author} {\bibfnamefont
  {J.}~\bibnamefont {Alonso}},\ and\ \bibinfo {author} {\bibfnamefont {J.~P.}\
  \bibnamefont {Home}},\ }\bibfield  {title} {\bibinfo {title} {Sustained
  state-independent quantum contextual correlations from a single ion},\ }\href
  {https://doi.org/10.1103/PhysRevLett.120.180401} {\bibfield  {journal}
  {\bibinfo  {journal} {Phys. Rev. Lett.}\ }\textbf {\bibinfo {volume} {120}},\
  \bibinfo {pages} {180401} (\bibinfo {year} {2018})}\BibitemShut {NoStop}%
\bibitem [{\citenamefont {Moussa}\ \emph {et~al.}(2010)\citenamefont {Moussa},
  \citenamefont {Ryan}, \citenamefont {Cory},\ and\ \citenamefont
  {Laflamme}}]{MRCL10}%
  \BibitemOpen
  \bibfield  {author} {\bibinfo {author} {\bibfnamefont {O.}~\bibnamefont
  {Moussa}}, \bibinfo {author} {\bibfnamefont {C.~A.}\ \bibnamefont {Ryan}},
  \bibinfo {author} {\bibfnamefont {D.~G.}\ \bibnamefont {Cory}},\ and\
  \bibinfo {author} {\bibfnamefont {R.}~\bibnamefont {Laflamme}},\ }\bibfield
  {title} {\bibinfo {title} {Testing contextuality on quantum ensembles with
  one clean qubit},\ }\href {https://doi.org/10.1103/PhysRevLett.104.160501}
  {\bibfield  {journal} {\bibinfo  {journal} {Phys. Rev. Lett.}\ }\textbf
  {\bibinfo {volume} {104}},\ \bibinfo {pages} {160501} (\bibinfo {year}
  {2010})}\BibitemShut {NoStop}%
\bibitem [{\citenamefont {Jerger}\ \emph {et~al.}(2016)\citenamefont {Jerger},
  \citenamefont {Reshitnyk}, \citenamefont {Oppliger}, \citenamefont
  {Potocnik}, \citenamefont {Mondal}, \citenamefont {Wallraff}, \citenamefont
  {Goodenough}, \citenamefont {Wehner}, \citenamefont {Juliusson},
  \citenamefont {Langford},\ and\ \citenamefont {Fedorov}}]{Jerger2016}%
  \BibitemOpen
  \bibfield  {author} {\bibinfo {author} {\bibfnamefont {M.}~\bibnamefont
  {Jerger}}, \bibinfo {author} {\bibfnamefont {Y.}~\bibnamefont {Reshitnyk}},
  \bibinfo {author} {\bibfnamefont {M.}~\bibnamefont {Oppliger}}, \bibinfo
  {author} {\bibfnamefont {A.}~\bibnamefont {Potocnik}}, \bibinfo {author}
  {\bibfnamefont {M.}~\bibnamefont {Mondal}}, \bibinfo {author} {\bibfnamefont
  {A.}~\bibnamefont {Wallraff}}, \bibinfo {author} {\bibfnamefont
  {K.}~\bibnamefont {Goodenough}}, \bibinfo {author} {\bibfnamefont
  {S.}~\bibnamefont {Wehner}}, \bibinfo {author} {\bibfnamefont
  {K.}~\bibnamefont {Juliusson}}, \bibinfo {author} {\bibfnamefont {N.~K.}\
  \bibnamefont {Langford}},\ and\ \bibinfo {author} {\bibfnamefont
  {A.}~\bibnamefont {Fedorov}},\ }\bibfield  {title} {\bibinfo {title}
  {Contextuality without nonlocality in a superconducting quantum system},\
  }\href {https://doi.org/10.1038/ncomms12930} {\bibfield  {journal} {\bibinfo
  {journal} {Nat. Commun.}\ }\textbf {\bibinfo {volume} {7}},\ \bibinfo {pages}
  {1} (\bibinfo {year} {2016})}\BibitemShut {NoStop}%
\bibitem [{\citenamefont {van Dam}\ \emph {et~al.}(2019)\citenamefont {van
  Dam}, \citenamefont {Cramer}, \citenamefont {Taminiau},\ and\ \citenamefont
  {Hanson}}]{VCTH19}%
  \BibitemOpen
  \bibfield  {author} {\bibinfo {author} {\bibfnamefont {S.~B.}\ \bibnamefont
  {van Dam}}, \bibinfo {author} {\bibfnamefont {J.}~\bibnamefont {Cramer}},
  \bibinfo {author} {\bibfnamefont {T.~H.}\ \bibnamefont {Taminiau}},\ and\
  \bibinfo {author} {\bibfnamefont {R.}~\bibnamefont {Hanson}},\ }\bibfield
  {title} {\bibinfo {title} {Multipartite entanglement generation and
  contextuality tests using nondestructive three-qubit parity measurements},\
  }\href {https://doi.org/10.1103/PhysRevLett.123.050401} {\bibfield  {journal}
  {\bibinfo  {journal} {Phys. Rev. Lett.}\ }\textbf {\bibinfo {volume} {123}},\
  \bibinfo {pages} {050401} (\bibinfo {year} {2019})}\BibitemShut {NoStop}%
\bibitem [{\citenamefont {Spekkens}(2005)}]{Spekkens2005}%
  \BibitemOpen
  \bibfield  {author} {\bibinfo {author} {\bibfnamefont {R.~W.}\ \bibnamefont
  {Spekkens}},\ }\bibfield  {title} {\bibinfo {title} {Contextuality for
  preparations, transformations, and unsharp measurements},\ }\href
  {https://doi.org/10.1103/PhysRevA.71.052108} {\bibfield  {journal} {\bibinfo
  {journal} {Phys. Rev. A}\ }\textbf {\bibinfo {volume} {71}},\ \bibinfo
  {pages} {052108} (\bibinfo {year} {2005})}\BibitemShut {NoStop}%
\bibitem [{\citenamefont {Spekkens}(2014)}]{Spekkens2014}%
  \BibitemOpen
  \bibfield  {author} {\bibinfo {author} {\bibfnamefont {R.~W.}\ \bibnamefont
  {Spekkens}},\ }\bibfield  {title} {\bibinfo {title} {The status of
  determinism in proofs of the impossibility of a noncontextual model of
  quantum theory},\ }\href {https://doi.org/10.1007/s10701-014-9833-x}
  {\bibfield  {journal} {\bibinfo  {journal} {Found. Phys.}\ }\textbf {\bibinfo
  {volume} {44}},\ \bibinfo {pages} {1125} (\bibinfo {year}
  {2014})}\BibitemShut {NoStop}%
\bibitem [{\citenamefont {{Chiribella}}\ and\ \citenamefont
  {{Yuan}}(2014)}]{YuanXiao2014}%
  \BibitemOpen
  \bibfield  {author} {\bibinfo {author} {\bibfnamefont {G.}~\bibnamefont
  {{Chiribella}}}\ and\ \bibinfo {author} {\bibfnamefont {X.}~\bibnamefont
  {{Yuan}}},\ }\href {https://ui.adsabs.harvard.edu/abs/2014arXiv1404.3348C}
  {\bibinfo {title} {Measurement sharpness cuts nonlocality and contextuality
  in every physical theory}} (\bibinfo {year} {2014}),\ \Eprint
  {https://arxiv.org/abs/1404.3348} {arXiv:1404.3348 [quant-ph]} \BibitemShut
  {NoStop}%
\bibitem [{\citenamefont {Cabello}(2019)}]{Cabello2019Quantumcorrelations}%
  \BibitemOpen
  \bibfield  {author} {\bibinfo {author} {\bibfnamefont {A.}~\bibnamefont
  {Cabello}},\ }\bibfield  {title} {\bibinfo {title} {Quantum correlations from
  simple assumptions},\ }\href {https://doi.org/10.1103/PhysRevA.100.032120}
  {\bibfield  {journal} {\bibinfo  {journal} {Phys. Rev. A}\ }\textbf {\bibinfo
  {volume} {100}},\ \bibinfo {pages} {032120} (\bibinfo {year}
  {2019})}\BibitemShut {NoStop}%
\bibitem [{\citenamefont {Heinosaari}\ \emph {et~al.}(2016)\citenamefont
  {Heinosaari}, \citenamefont {Miyadera},\ and\ \citenamefont
  {Ziman}}]{Heinosaari2016}%
  \BibitemOpen
  \bibfield  {author} {\bibinfo {author} {\bibfnamefont {T.}~\bibnamefont
  {Heinosaari}}, \bibinfo {author} {\bibfnamefont {T.}~\bibnamefont
  {Miyadera}},\ and\ \bibinfo {author} {\bibfnamefont {M.}~\bibnamefont
  {Ziman}},\ }\bibfield  {title} {\bibinfo {title} {An invitation to quantum
  incompatibility},\ }\href {https://doi.org/10.1088/1751-8113/49/12/123001}
  {\bibfield  {journal} {\bibinfo  {journal} {J. Phys. A: Math. Theor.}\
  }\textbf {\bibinfo {volume} {49}},\ \bibinfo {pages} {123001} (\bibinfo
  {year} {2016})}\BibitemShut {NoStop}%
\bibitem [{\citenamefont {Pearle}(1970)}]{Pearle1970}%
  \BibitemOpen
  \bibfield  {author} {\bibinfo {author} {\bibfnamefont {P.~M.}\ \bibnamefont
  {Pearle}},\ }\bibfield  {title} {\bibinfo {title} {Hidden-variable example
  based upon data rejection},\ }\href {https://doi.org/10.1103/PhysRevD.2.1418}
  {\bibfield  {journal} {\bibinfo  {journal} {Phys. Rev. D}\ }\textbf {\bibinfo
  {volume} {2}},\ \bibinfo {pages} {1418} (\bibinfo {year} {1970})}\BibitemShut
  {NoStop}%
\bibitem [{\citenamefont {Garg}\ and\ \citenamefont
  {Mermin}(1987)}]{GargMermin1987}%
  \BibitemOpen
  \bibfield  {author} {\bibinfo {author} {\bibfnamefont {A.}~\bibnamefont
  {Garg}}\ and\ \bibinfo {author} {\bibfnamefont {N.~D.}\ \bibnamefont
  {Mermin}},\ }\bibfield  {title} {\bibinfo {title} {Detector inefficiencies in
  the einstein-podolsky-rosen experiment},\ }\href
  {https://doi.org/10.1103/PhysRevD.35.3831} {\bibfield  {journal} {\bibinfo
  {journal} {Phys. Rev. D}\ }\textbf {\bibinfo {volume} {35}},\ \bibinfo
  {pages} {3831} (\bibinfo {year} {1987})}\BibitemShut {NoStop}%
\bibitem [{\citenamefont {Hensen}\ \emph {et~al.}(2015)\citenamefont {Hensen},
  \citenamefont {Bernien}, \citenamefont {Dr{\'e}au}, \citenamefont {Reiserer},
  \citenamefont {Kalb}, \citenamefont {Blok}, \citenamefont {Ruitenberg},
  \citenamefont {Vermeulen}, \citenamefont {Schouten}, \citenamefont
  {Abell{\'a}n}, \citenamefont {Amaya}, \citenamefont {Pruneri}, \citenamefont
  {Mitchell}, \citenamefont {Markham}, \citenamefont {Twitchen}, \citenamefont
  {Elkouss}, \citenamefont {Wehner}, \citenamefont {Taminiau},\ and\
  \citenamefont {Hanson}}]{HBD15}%
  \BibitemOpen
  \bibfield  {author} {\bibinfo {author} {\bibfnamefont {B.}~\bibnamefont
  {Hensen}}, \bibinfo {author} {\bibfnamefont {H.}~\bibnamefont {Bernien}},
  \bibinfo {author} {\bibfnamefont {A.~E.}\ \bibnamefont {Dr{\'e}au}}, \bibinfo
  {author} {\bibfnamefont {A.}~\bibnamefont {Reiserer}}, \bibinfo {author}
  {\bibfnamefont {N.}~\bibnamefont {Kalb}}, \bibinfo {author} {\bibfnamefont
  {M.~S.}\ \bibnamefont {Blok}}, \bibinfo {author} {\bibfnamefont
  {J.}~\bibnamefont {Ruitenberg}}, \bibinfo {author} {\bibfnamefont {R.~F.}\
  \bibnamefont {Vermeulen}}, \bibinfo {author} {\bibfnamefont {R.~N.}\
  \bibnamefont {Schouten}}, \bibinfo {author} {\bibfnamefont {C.}~\bibnamefont
  {Abell{\'a}n}}, \bibinfo {author} {\bibfnamefont {W.}~\bibnamefont {Amaya}},
  \bibinfo {author} {\bibfnamefont {V.}~\bibnamefont {Pruneri}}, \bibinfo
  {author} {\bibfnamefont {M.~W.}\ \bibnamefont {Mitchell}}, \bibinfo {author}
  {\bibfnamefont {M.}~\bibnamefont {Markham}}, \bibinfo {author} {\bibfnamefont
  {D.~J.}\ \bibnamefont {Twitchen}}, \bibinfo {author} {\bibfnamefont
  {D.}~\bibnamefont {Elkouss}}, \bibinfo {author} {\bibfnamefont
  {S.}~\bibnamefont {Wehner}}, \bibinfo {author} {\bibfnamefont {T.~H.}\
  \bibnamefont {Taminiau}},\ and\ \bibinfo {author} {\bibfnamefont
  {R.}~\bibnamefont {Hanson}},\ }\bibfield  {title} {\bibinfo {title}
  {Loophole-free bell inequality violation using electron spins separated by
  1.3 kilometres},\ }\href {https://doi.org/10.1038/nature15759} {\bibfield
  {journal} {\bibinfo  {journal} {Nature}\ }\textbf {\bibinfo {volume} {526}},\
  \bibinfo {pages} {682} (\bibinfo {year} {2015})}\BibitemShut {NoStop}%
\bibitem [{\citenamefont {Giustina}\ \emph {et~al.}(2015)\citenamefont
  {Giustina}, \citenamefont {Versteegh}, \citenamefont {Wengerowsky},
  \citenamefont {Handsteiner}, \citenamefont {Hochrainer}, \citenamefont
  {Phelan}, \citenamefont {Steinlechner}, \citenamefont {Kofler}, \citenamefont
  {Larsson}, \citenamefont {Abell\'an}, \citenamefont {Amaya}, \citenamefont
  {Pruneri}, \citenamefont {Mitchell}, \citenamefont {Beyer}, \citenamefont
  {Gerrits}, \citenamefont {Lita}, \citenamefont {Shalm}, \citenamefont {Nam},
  \citenamefont {Scheidl}, \citenamefont {Ursin}, \citenamefont {Wittmann},\
  and\ \citenamefont {Zeilinger}}]{GVW15}%
  \BibitemOpen
  \bibfield  {author} {\bibinfo {author} {\bibfnamefont {M.}~\bibnamefont
  {Giustina}}, \bibinfo {author} {\bibfnamefont {M.~A.~M.}\ \bibnamefont
  {Versteegh}}, \bibinfo {author} {\bibfnamefont {S.}~\bibnamefont
  {Wengerowsky}}, \bibinfo {author} {\bibfnamefont {J.}~\bibnamefont
  {Handsteiner}}, \bibinfo {author} {\bibfnamefont {A.}~\bibnamefont
  {Hochrainer}}, \bibinfo {author} {\bibfnamefont {K.}~\bibnamefont {Phelan}},
  \bibinfo {author} {\bibfnamefont {F.}~\bibnamefont {Steinlechner}}, \bibinfo
  {author} {\bibfnamefont {J.}~\bibnamefont {Kofler}}, \bibinfo {author}
  {\bibfnamefont {J.-A.}\ \bibnamefont {Larsson}}, \bibinfo {author}
  {\bibfnamefont {C.}~\bibnamefont {Abell\'an}}, \bibinfo {author}
  {\bibfnamefont {W.}~\bibnamefont {Amaya}}, \bibinfo {author} {\bibfnamefont
  {V.}~\bibnamefont {Pruneri}}, \bibinfo {author} {\bibfnamefont {M.~W.}\
  \bibnamefont {Mitchell}}, \bibinfo {author} {\bibfnamefont {J.}~\bibnamefont
  {Beyer}}, \bibinfo {author} {\bibfnamefont {T.}~\bibnamefont {Gerrits}},
  \bibinfo {author} {\bibfnamefont {A.~E.}\ \bibnamefont {Lita}}, \bibinfo
  {author} {\bibfnamefont {L.~K.}\ \bibnamefont {Shalm}}, \bibinfo {author}
  {\bibfnamefont {S.~W.}\ \bibnamefont {Nam}}, \bibinfo {author} {\bibfnamefont
  {T.}~\bibnamefont {Scheidl}}, \bibinfo {author} {\bibfnamefont
  {R.}~\bibnamefont {Ursin}}, \bibinfo {author} {\bibfnamefont
  {B.}~\bibnamefont {Wittmann}},\ and\ \bibinfo {author} {\bibfnamefont
  {A.}~\bibnamefont {Zeilinger}},\ }\bibfield  {title} {\bibinfo {title}
  {Significant-loophole-free test of bell's theorem with entangled photons},\
  }\href {https://doi.org/10.1103/PhysRevLett.115.250401} {\bibfield  {journal}
  {\bibinfo  {journal} {Phys. Rev. Lett.}\ }\textbf {\bibinfo {volume} {115}},\
  \bibinfo {pages} {250401} (\bibinfo {year} {2015})}\BibitemShut {NoStop}%
\bibitem [{\citenamefont {Shalm}\ \emph {et~al.}(2015)\citenamefont {Shalm},
  \citenamefont {Meyer-Scott}, \citenamefont {Christensen}, \citenamefont
  {Bierhorst}, \citenamefont {Wayne}, \citenamefont {Stevens}, \citenamefont
  {Gerrits}, \citenamefont {Glancy}, \citenamefont {Hamel}, \citenamefont
  {Allman}, \citenamefont {Coakley}, \citenamefont {Dyer}, \citenamefont
  {Hodge}, \citenamefont {Lita}, \citenamefont {Verma}, \citenamefont
  {Lambrocco}, \citenamefont {Tortorici}, \citenamefont {Migdall},
  \citenamefont {Zhang}, \citenamefont {Kumor}, \citenamefont {Farr},
  \citenamefont {Marsili}, \citenamefont {Shaw}, \citenamefont {Stern},
  \citenamefont {Abell\'an}, \citenamefont {Amaya}, \citenamefont {Pruneri},
  \citenamefont {Jennewein}, \citenamefont {Mitchell}, \citenamefont {Kwiat},
  \citenamefont {Bienfang}, \citenamefont {Mirin}, \citenamefont {Knill},\ and\
  \citenamefont {Nam}}]{SMC15}%
  \BibitemOpen
  \bibfield  {author} {\bibinfo {author} {\bibfnamefont {L.~K.}\ \bibnamefont
  {Shalm}}, \bibinfo {author} {\bibfnamefont {E.}~\bibnamefont {Meyer-Scott}},
  \bibinfo {author} {\bibfnamefont {B.~G.}\ \bibnamefont {Christensen}},
  \bibinfo {author} {\bibfnamefont {P.}~\bibnamefont {Bierhorst}}, \bibinfo
  {author} {\bibfnamefont {M.~A.}\ \bibnamefont {Wayne}}, \bibinfo {author}
  {\bibfnamefont {M.~J.}\ \bibnamefont {Stevens}}, \bibinfo {author}
  {\bibfnamefont {T.}~\bibnamefont {Gerrits}}, \bibinfo {author} {\bibfnamefont
  {S.}~\bibnamefont {Glancy}}, \bibinfo {author} {\bibfnamefont {D.~R.}\
  \bibnamefont {Hamel}}, \bibinfo {author} {\bibfnamefont {M.~S.}\ \bibnamefont
  {Allman}}, \bibinfo {author} {\bibfnamefont {K.~J.}\ \bibnamefont {Coakley}},
  \bibinfo {author} {\bibfnamefont {S.~D.}\ \bibnamefont {Dyer}}, \bibinfo
  {author} {\bibfnamefont {C.}~\bibnamefont {Hodge}}, \bibinfo {author}
  {\bibfnamefont {A.~E.}\ \bibnamefont {Lita}}, \bibinfo {author}
  {\bibfnamefont {V.~B.}\ \bibnamefont {Verma}}, \bibinfo {author}
  {\bibfnamefont {C.}~\bibnamefont {Lambrocco}}, \bibinfo {author}
  {\bibfnamefont {E.}~\bibnamefont {Tortorici}}, \bibinfo {author}
  {\bibfnamefont {A.~L.}\ \bibnamefont {Migdall}}, \bibinfo {author}
  {\bibfnamefont {Y.}~\bibnamefont {Zhang}}, \bibinfo {author} {\bibfnamefont
  {D.~R.}\ \bibnamefont {Kumor}}, \bibinfo {author} {\bibfnamefont {W.~H.}\
  \bibnamefont {Farr}}, \bibinfo {author} {\bibfnamefont {F.}~\bibnamefont
  {Marsili}}, \bibinfo {author} {\bibfnamefont {M.~D.}\ \bibnamefont {Shaw}},
  \bibinfo {author} {\bibfnamefont {J.~A.}\ \bibnamefont {Stern}}, \bibinfo
  {author} {\bibfnamefont {C.}~\bibnamefont {Abell\'an}}, \bibinfo {author}
  {\bibfnamefont {W.}~\bibnamefont {Amaya}}, \bibinfo {author} {\bibfnamefont
  {V.}~\bibnamefont {Pruneri}}, \bibinfo {author} {\bibfnamefont
  {T.}~\bibnamefont {Jennewein}}, \bibinfo {author} {\bibfnamefont {M.~W.}\
  \bibnamefont {Mitchell}}, \bibinfo {author} {\bibfnamefont {P.~G.}\
  \bibnamefont {Kwiat}}, \bibinfo {author} {\bibfnamefont {J.~C.}\ \bibnamefont
  {Bienfang}}, \bibinfo {author} {\bibfnamefont {R.~P.}\ \bibnamefont {Mirin}},
  \bibinfo {author} {\bibfnamefont {E.}~\bibnamefont {Knill}},\ and\ \bibinfo
  {author} {\bibfnamefont {S.~W.}\ \bibnamefont {Nam}},\ }\bibfield  {title}
  {\bibinfo {title} {Strong loophole-free test of local realism},\ }\href
  {https://doi.org/10.1103/PhysRevLett.115.250402} {\bibfield  {journal}
  {\bibinfo  {journal} {Phys. Rev. Lett.}\ }\textbf {\bibinfo {volume} {115}},\
  \bibinfo {pages} {250402} (\bibinfo {year} {2015})}\BibitemShut {NoStop}%
\bibitem [{\citenamefont {Rosenfeld}\ \emph {et~al.}(2017)\citenamefont
  {Rosenfeld}, \citenamefont {Burchardt}, \citenamefont {Garthoff},
  \citenamefont {Redeker}, \citenamefont {Ortegel}, \citenamefont {Rau},\ and\
  \citenamefont {Weinfurter}}]{RBG17}%
  \BibitemOpen
  \bibfield  {author} {\bibinfo {author} {\bibfnamefont {W.}~\bibnamefont
  {Rosenfeld}}, \bibinfo {author} {\bibfnamefont {D.}~\bibnamefont
  {Burchardt}}, \bibinfo {author} {\bibfnamefont {R.}~\bibnamefont {Garthoff}},
  \bibinfo {author} {\bibfnamefont {K.}~\bibnamefont {Redeker}}, \bibinfo
  {author} {\bibfnamefont {N.}~\bibnamefont {Ortegel}}, \bibinfo {author}
  {\bibfnamefont {M.}~\bibnamefont {Rau}},\ and\ \bibinfo {author}
  {\bibfnamefont {H.}~\bibnamefont {Weinfurter}},\ }\bibfield  {title}
  {\bibinfo {title} {Event-ready bell test using entangled atoms simultaneously
  closing detection and locality loopholes},\ }\href
  {https://doi.org/10.1103/PhysRevLett.119.010402} {\bibfield  {journal}
  {\bibinfo  {journal} {Phys. Rev. Lett.}\ }\textbf {\bibinfo {volume} {119}},\
  \bibinfo {pages} {010402} (\bibinfo {year} {2017})}\BibitemShut {NoStop}%
\bibitem [{\citenamefont {G\"uhne}\ \emph {et~al.}(2010)\citenamefont
  {G\"uhne}, \citenamefont {Kleinmann}, \citenamefont {Cabello}, \citenamefont
  {Larsson}, \citenamefont {Kirchmair}, \citenamefont {Z\"ahringer},
  \citenamefont {Gerritsma},\ and\ \citenamefont {Roos}}]{Guhne&Cabello2010}%
  \BibitemOpen
  \bibfield  {author} {\bibinfo {author} {\bibfnamefont {O.}~\bibnamefont
  {G\"uhne}}, \bibinfo {author} {\bibfnamefont {M.}~\bibnamefont {Kleinmann}},
  \bibinfo {author} {\bibfnamefont {A.}~\bibnamefont {Cabello}}, \bibinfo
  {author} {\bibfnamefont {J.-A.}\ \bibnamefont {Larsson}}, \bibinfo {author}
  {\bibfnamefont {G.}~\bibnamefont {Kirchmair}}, \bibinfo {author}
  {\bibfnamefont {F.}~\bibnamefont {Z\"ahringer}}, \bibinfo {author}
  {\bibfnamefont {R.}~\bibnamefont {Gerritsma}},\ and\ \bibinfo {author}
  {\bibfnamefont {C.~F.}\ \bibnamefont {Roos}},\ }\bibfield  {title} {\bibinfo
  {title} {Compatibility and noncontextuality for sequential measurements},\
  }\href {https://doi.org/10.1103/PhysRevA.81.022121} {\bibfield  {journal}
  {\bibinfo  {journal} {Phys. Rev. A}\ }\textbf {\bibinfo {volume} {81}},\
  \bibinfo {pages} {022121} (\bibinfo {year} {2010})}\BibitemShut {NoStop}%
\bibitem [{\citenamefont {Szangolies}\ \emph {et~al.}(2013)\citenamefont
  {Szangolies}, \citenamefont {Kleinmann},\ and\ \citenamefont
  {G\"uhne}}]{Szangolies2013Tests}%
  \BibitemOpen
  \bibfield  {author} {\bibinfo {author} {\bibfnamefont {J.}~\bibnamefont
  {Szangolies}}, \bibinfo {author} {\bibfnamefont {M.}~\bibnamefont
  {Kleinmann}},\ and\ \bibinfo {author} {\bibfnamefont {O.}~\bibnamefont
  {G\"uhne}},\ }\bibfield  {title} {\bibinfo {title} {Tests against
  noncontextual models with measurement disturbances},\ }\href
  {https://doi.org/10.1103/PhysRevA.87.050101} {\bibfield  {journal} {\bibinfo
  {journal} {Phys. Rev. A}\ }\textbf {\bibinfo {volume} {87}},\ \bibinfo
  {pages} {050101} (\bibinfo {year} {2013})}\BibitemShut {NoStop}%
\bibitem [{\citenamefont {Cabello}\ and\ \citenamefont
  {Cunha}(2011)}]{Cabello2011qutrit}%
  \BibitemOpen
  \bibfield  {author} {\bibinfo {author} {\bibfnamefont {A.}~\bibnamefont
  {Cabello}}\ and\ \bibinfo {author} {\bibfnamefont {M.~T.}\ \bibnamefont
  {Cunha}},\ }\bibfield  {title} {\bibinfo {title} {Proposal of a two-qutrit
  contextuality test free of the finite precision and compatibility
  loopholes},\ }\href {https://doi.org/10.1103/PhysRevLett.106.190401}
  {\bibfield  {journal} {\bibinfo  {journal} {Phys. Rev. Lett.}\ }\textbf
  {\bibinfo {volume} {106}},\ \bibinfo {pages} {190401} (\bibinfo {year}
  {2011})}\BibitemShut {NoStop}%
\bibitem [{\citenamefont {Hu}\ \emph {et~al.}(2016)\citenamefont {Hu},
  \citenamefont {Chen}, \citenamefont {Liu}, \citenamefont {Guo}, \citenamefont
  {Huang}, \citenamefont {Zhou}, \citenamefont {Han}, \citenamefont {Li},\ and\
  \citenamefont {Guo}}]{HuXiaomin2016}%
  \BibitemOpen
  \bibfield  {author} {\bibinfo {author} {\bibfnamefont {X.~M.}\ \bibnamefont
  {Hu}}, \bibinfo {author} {\bibfnamefont {J.~S.}\ \bibnamefont {Chen}},
  \bibinfo {author} {\bibfnamefont {B.~H.}\ \bibnamefont {Liu}}, \bibinfo
  {author} {\bibfnamefont {Y.}~\bibnamefont {Guo}}, \bibinfo {author}
  {\bibfnamefont {Y.~F.}\ \bibnamefont {Huang}}, \bibinfo {author}
  {\bibfnamefont {Z.~Q.}\ \bibnamefont {Zhou}}, \bibinfo {author}
  {\bibfnamefont {Y.~J.}\ \bibnamefont {Han}}, \bibinfo {author} {\bibfnamefont
  {C.~F.}\ \bibnamefont {Li}},\ and\ \bibinfo {author} {\bibfnamefont {G.~C.}\
  \bibnamefont {Guo}},\ }\bibfield  {title} {\bibinfo {title} {Experimental
  test of compatibility-loophole-free contextuality with spatially separated
  entangled qutrits},\ }\href {https://doi.org/10.1103/PhysRevLett.117.170403}
  {\bibfield  {journal} {\bibinfo  {journal} {Phys. Rev. Lett.}\ }\textbf
  {\bibinfo {volume} {117}},\ \bibinfo {pages} {170403} (\bibinfo {year}
  {2016})}\BibitemShut {NoStop}%
\bibitem [{\citenamefont {Malinowski}\ \emph {et~al.}(2018)\citenamefont
  {Malinowski}, \citenamefont {Zhang}, \citenamefont {Leupold}, \citenamefont
  {Cabello}, \citenamefont {Alonso},\ and\ \citenamefont
  {Home}}]{Malinowski&HomeJP&Cabello2018}%
  \BibitemOpen
  \bibfield  {author} {\bibinfo {author} {\bibfnamefont {M.}~\bibnamefont
  {Malinowski}}, \bibinfo {author} {\bibfnamefont {C.}~\bibnamefont {Zhang}},
  \bibinfo {author} {\bibfnamefont {F.~M.}\ \bibnamefont {Leupold}}, \bibinfo
  {author} {\bibfnamefont {A.}~\bibnamefont {Cabello}}, \bibinfo {author}
  {\bibfnamefont {J.}~\bibnamefont {Alonso}},\ and\ \bibinfo {author}
  {\bibfnamefont {J.~P.}\ \bibnamefont {Home}},\ }\bibfield  {title} {\bibinfo
  {title} {Probing the limits of correlations in an indivisible quantum
  system},\ }\href {https://doi.org/10.1103/PhysRevA.98.050102} {\bibfield
  {journal} {\bibinfo  {journal} {Phys. Rev. A}\ }\textbf {\bibinfo {volume}
  {98}},\ \bibinfo {pages} {050102} (\bibinfo {year} {2018})}\BibitemShut
  {NoStop}%
\bibitem [{\citenamefont {Home}(2013)}]{home2013quantum}%
  \BibitemOpen
  \bibfield  {author} {\bibinfo {author} {\bibfnamefont {J.~P.}\ \bibnamefont
  {Home}},\ }\bibfield  {title} {\bibinfo {title} {Quantum science and
  metrology with mixed-species ion chains},\ }\href@noop {} {\bibfield
  {journal} {\bibinfo  {journal} {Adv. At. Mol. Opt. Phys.}\ }\textbf {\bibinfo
  {volume} {62}},\ \bibinfo {pages} {231} (\bibinfo {year} {2013})}\BibitemShut
  {NoStop}%
\bibitem [{\citenamefont {Tan}\ \emph {et~al.}(2015)\citenamefont {Tan},
  \citenamefont {Gaebler}, \citenamefont {Lin}, \citenamefont {Wan},
  \citenamefont {Bowler}, \citenamefont {Leibfried},\ and\ \citenamefont
  {Wineland}}]{Tan&Wineland2015}%
  \BibitemOpen
  \bibfield  {author} {\bibinfo {author} {\bibfnamefont {T.~R.}\ \bibnamefont
  {Tan}}, \bibinfo {author} {\bibfnamefont {J.~P.}\ \bibnamefont {Gaebler}},
  \bibinfo {author} {\bibfnamefont {Y.}~\bibnamefont {Lin}}, \bibinfo {author}
  {\bibfnamefont {Y.}~\bibnamefont {Wan}}, \bibinfo {author} {\bibfnamefont
  {R.}~\bibnamefont {Bowler}}, \bibinfo {author} {\bibfnamefont
  {D.}~\bibnamefont {Leibfried}},\ and\ \bibinfo {author} {\bibfnamefont
  {D.~J.}\ \bibnamefont {Wineland}},\ }\bibfield  {title} {\bibinfo {title}
  {Multi-element logic gates for trapped-ion qubits},\ }\href
  {https://doi.org/10.1038/nature16186} {\bibfield  {journal} {\bibinfo
  {journal} {Nature}\ }\textbf {\bibinfo {volume} {528}},\ \bibinfo {pages}
  {380} (\bibinfo {year} {2015})}\BibitemShut {NoStop}%
\bibitem [{\citenamefont {Ballance}\ \emph {et~al.}(2015)\citenamefont
  {Ballance}, \citenamefont {Schafer}, \citenamefont {Home}, \citenamefont
  {Szwer}, \citenamefont {Webster}, \citenamefont {Allcock}, \citenamefont
  {Linke}, \citenamefont {Harty}, \citenamefont {Aude~Craik}, \citenamefont
  {Stacey}, \citenamefont {Steane},\ and\ \citenamefont
  {Lucas}}]{Ballance2015}%
  \BibitemOpen
  \bibfield  {author} {\bibinfo {author} {\bibfnamefont {C.~J.}\ \bibnamefont
  {Ballance}}, \bibinfo {author} {\bibfnamefont {V.~M.}\ \bibnamefont
  {Schafer}}, \bibinfo {author} {\bibfnamefont {J.~P.}\ \bibnamefont {Home}},
  \bibinfo {author} {\bibfnamefont {D.~J.}\ \bibnamefont {Szwer}}, \bibinfo
  {author} {\bibfnamefont {S.~C.}\ \bibnamefont {Webster}}, \bibinfo {author}
  {\bibfnamefont {D.~T.}\ \bibnamefont {Allcock}}, \bibinfo {author}
  {\bibfnamefont {N.~M.}\ \bibnamefont {Linke}}, \bibinfo {author}
  {\bibfnamefont {T.~P.}\ \bibnamefont {Harty}}, \bibinfo {author}
  {\bibfnamefont {D.~P.}\ \bibnamefont {Aude~Craik}}, \bibinfo {author}
  {\bibfnamefont {D.~N.}\ \bibnamefont {Stacey}}, \bibinfo {author}
  {\bibfnamefont {A.~M.}\ \bibnamefont {Steane}},\ and\ \bibinfo {author}
  {\bibfnamefont {D.~M.}\ \bibnamefont {Lucas}},\ }\bibfield  {title} {\bibinfo
  {title} {Hybrid quantum logic and a test of bell's inequality using two
  different atomic isotopes},\ }\href {https://doi.org/10.1038/nature16184}
  {\bibfield  {journal} {\bibinfo  {journal} {Nature}\ }\textbf {\bibinfo
  {volume} {528}},\ \bibinfo {pages} {384} (\bibinfo {year}
  {2015})}\BibitemShut {NoStop}%
\bibitem [{\citenamefont {Inlek}\ \emph {et~al.}(2017)\citenamefont {Inlek},
  \citenamefont {Crocker}, \citenamefont {Lichtman}, \citenamefont {Sosnova},\
  and\ \citenamefont {Monroe}}]{Inlek2017}%
  \BibitemOpen
  \bibfield  {author} {\bibinfo {author} {\bibfnamefont {I.~V.}\ \bibnamefont
  {Inlek}}, \bibinfo {author} {\bibfnamefont {C.}~\bibnamefont {Crocker}},
  \bibinfo {author} {\bibfnamefont {M.}~\bibnamefont {Lichtman}}, \bibinfo
  {author} {\bibfnamefont {K.}~\bibnamefont {Sosnova}},\ and\ \bibinfo {author}
  {\bibfnamefont {C.}~\bibnamefont {Monroe}},\ }\bibfield  {title} {\bibinfo
  {title} {Multispecies trapped-ion node for quantum networking},\ }\href
  {https://doi.org/10.1103/PhysRevLett.118.250502} {\bibfield  {journal}
  {\bibinfo  {journal} {Phys. Rev. Lett.}\ }\textbf {\bibinfo {volume} {118}},\
  \bibinfo {pages} {250502} (\bibinfo {year} {2017})}\BibitemShut {NoStop}%
\bibitem [{\citenamefont {Negnevitsky}\ \emph {et~al.}(2018)\citenamefont
  {Negnevitsky}, \citenamefont {Marinelli}, \citenamefont {Mehta},
  \citenamefont {Lo}, \citenamefont {Flühmann},\ and\ \citenamefont
  {Home}}]{Home2018}%
  \BibitemOpen
  \bibfield  {author} {\bibinfo {author} {\bibfnamefont {V.}~\bibnamefont
  {Negnevitsky}}, \bibinfo {author} {\bibfnamefont {M.}~\bibnamefont
  {Marinelli}}, \bibinfo {author} {\bibfnamefont {K.~K.}\ \bibnamefont
  {Mehta}}, \bibinfo {author} {\bibfnamefont {H.~Y.}\ \bibnamefont {Lo}},
  \bibinfo {author} {\bibfnamefont {C.}~\bibnamefont {Flühmann}},\ and\
  \bibinfo {author} {\bibfnamefont {J.~P.}\ \bibnamefont {Home}},\ }\bibfield
  {title} {\bibinfo {title} {Repeated multi-qubit readout and feedback with a
  mixed-species trapped-ion register},\ }\href
  {https://doi.org/10.1038/s41586-018-0668-z} {\bibfield  {journal} {\bibinfo
  {journal} {Nature}\ }\textbf {\bibinfo {volume} {563}},\ \bibinfo {pages}
  {527} (\bibinfo {year} {2018})}\BibitemShut {NoStop}%
\bibitem [{\citenamefont {Bruzewicz}\ \emph {et~al.}(2019)\citenamefont
  {Bruzewicz}, \citenamefont {McConnell}, \citenamefont {Stuart}, \citenamefont
  {Sage},\ and\ \citenamefont {Chiaverini}}]{MIT2019}%
  \BibitemOpen
  \bibfield  {author} {\bibinfo {author} {\bibfnamefont {C.~D.}\ \bibnamefont
  {Bruzewicz}}, \bibinfo {author} {\bibfnamefont {R.}~\bibnamefont
  {McConnell}}, \bibinfo {author} {\bibfnamefont {J.}~\bibnamefont {Stuart}},
  \bibinfo {author} {\bibfnamefont {J.~M.}\ \bibnamefont {Sage}},\ and\
  \bibinfo {author} {\bibfnamefont {J.}~\bibnamefont {Chiaverini}},\ }\bibfield
   {title} {\bibinfo {title} {Dual-species, multi-qubit logic primitives for
  ca+/sr+ trapped-ion crystals},\ }\href
  {https://doi.org/10.1038/s41534-019-0218-z} {\bibfield  {journal} {\bibinfo
  {journal} {NPJ Quantum Inf.}\ }\textbf {\bibinfo {volume} {5}},\ \bibinfo
  {pages} {1} (\bibinfo {year} {2019})}\BibitemShut {NoStop}%
\bibitem [{\citenamefont {Isham}(2001)}]{Isham2001}%
  \BibitemOpen
  \bibfield  {author} {\bibinfo {author} {\bibfnamefont {C.~J.}\ \bibnamefont
  {Isham}},\ }\href@noop {} {\emph {\bibinfo {title} {Lectures on Quantum
  Theory}}}\ (\bibinfo  {publisher} {Imperial College Press},\ \bibinfo
  {address} {London},\ \bibinfo {year} {2001})\ p.\ \bibinfo {pages}
  {179}\BibitemShut {NoStop}%
\bibitem [{\citenamefont {Ara\'ujo}\ \emph {et~al.}(2013)\citenamefont
  {Ara\'ujo}, \citenamefont {Quintino}, \citenamefont {Budroni}, \citenamefont
  {Cunha},\ and\ \citenamefont {Cabello}}]{Araujo&Cabello2013}%
  \BibitemOpen
  \bibfield  {author} {\bibinfo {author} {\bibfnamefont {M.}~\bibnamefont
  {Ara\'ujo}}, \bibinfo {author} {\bibfnamefont {M.~T.}\ \bibnamefont
  {Quintino}}, \bibinfo {author} {\bibfnamefont {C.}~\bibnamefont {Budroni}},
  \bibinfo {author} {\bibfnamefont {M.~T.}\ \bibnamefont {Cunha}},\ and\
  \bibinfo {author} {\bibfnamefont {A.}~\bibnamefont {Cabello}},\ }\bibfield
  {title} {\bibinfo {title} {All noncontextuality inequalities for the
  $n$-cycle scenario},\ }\href {https://doi.org/10.1103/PhysRevA.88.022118}
  {\bibfield  {journal} {\bibinfo  {journal} {Phys. Rev. A}\ }\textbf {\bibinfo
  {volume} {88}},\ \bibinfo {pages} {022118} (\bibinfo {year}
  {2013})}\BibitemShut {NoStop}%
\bibitem [{\citenamefont {Vorob’ev}(1962)}]{Vorob'yev62}%
  \BibitemOpen
  \bibfield  {author} {\bibinfo {author} {\bibfnamefont {N.~N.}\ \bibnamefont
  {Vorob’ev}},\ }\bibfield  {title} {\bibinfo {title} {Consistent families of
  measures and their extensions},\ }\href@noop {} {\bibfield  {journal}
  {\bibinfo  {journal} {Theory of Probability \& Its Applications}\ }\textbf
  {\bibinfo {volume} {7}},\ \bibinfo {pages} {147} (\bibinfo {year}
  {1962})}\BibitemShut {NoStop}%
\bibitem [{\citenamefont {Vorob’ev}(1963)}]{Vorob'yev63}%
  \BibitemOpen
  \bibfield  {author} {\bibinfo {author} {\bibfnamefont {N.}~\bibnamefont
  {Vorob’ev}},\ }\bibfield  {title} {\bibinfo {title} {Markov measures and
  markov extensions},\ }\href@noop {} {\bibfield  {journal} {\bibinfo
  {journal} {Theory of Probability \& Its Applications}\ }\textbf {\bibinfo
  {volume} {8}},\ \bibinfo {pages} {420} (\bibinfo {year} {1963})}\BibitemShut
  {NoStop}%
\bibitem [{\citenamefont {Clauser}\ \emph {et~al.}(1969)\citenamefont
  {Clauser}, \citenamefont {Horne}, \citenamefont {Shimony},\ and\
  \citenamefont {Holt}}]{Clauser1969}%
  \BibitemOpen
  \bibfield  {author} {\bibinfo {author} {\bibfnamefont {J.~F.}\ \bibnamefont
  {Clauser}}, \bibinfo {author} {\bibfnamefont {M.~A.}\ \bibnamefont {Horne}},
  \bibinfo {author} {\bibfnamefont {A.}~\bibnamefont {Shimony}},\ and\ \bibinfo
  {author} {\bibfnamefont {R.~A.}\ \bibnamefont {Holt}},\ }\bibfield  {title}
  {\bibinfo {title} {Proposed experiment to test local hidden-variable
  theories},\ }\href {https://doi.org/10.1103/PhysRevLett.23.880} {\bibfield
  {journal} {\bibinfo  {journal} {Phys. Rev. Lett.}\ }\textbf {\bibinfo
  {volume} {23}},\ \bibinfo {pages} {880} (\bibinfo {year} {1969})}\BibitemShut
  {NoStop}%
\bibitem [{\citenamefont {Kleinmann}\ \emph {et~al.}(2012)\citenamefont
  {Kleinmann}, \citenamefont {Budroni}, \citenamefont {Larsson}, \citenamefont
  {G{\"u}hne},\ and\ \citenamefont {Cabello}}]{kleinmann2012optimal}%
  \BibitemOpen
  \bibfield  {author} {\bibinfo {author} {\bibfnamefont {M.}~\bibnamefont
  {Kleinmann}}, \bibinfo {author} {\bibfnamefont {C.}~\bibnamefont {Budroni}},
  \bibinfo {author} {\bibfnamefont {J.-{\AA}.}\ \bibnamefont {Larsson}},
  \bibinfo {author} {\bibfnamefont {O.}~\bibnamefont {G{\"u}hne}},\ and\
  \bibinfo {author} {\bibfnamefont {A.}~\bibnamefont {Cabello}},\ }\bibfield
  {title} {\bibinfo {title} {Optimal inequalities for state-independent
  contextuality},\ }\href@noop {} {\bibfield  {journal} {\bibinfo  {journal}
  {Phys. Rev. Lett.}\ }\textbf {\bibinfo {volume} {109}},\ \bibinfo {pages}
  {250402} (\bibinfo {year} {2012})}\BibitemShut {NoStop}%
\bibitem [{\citenamefont {Paul}(1990)}]{WolfgangRMP}%
  \BibitemOpen
  \bibfield  {author} {\bibinfo {author} {\bibfnamefont {W.}~\bibnamefont
  {Paul}},\ }\bibfield  {title} {\bibinfo {title} {Electromagnetic traps for
  charged and neutral particles},\ }\href
  {https://doi.org/10.1103/RevModPhys.62.531} {\bibfield  {journal} {\bibinfo
  {journal} {Rev. Mod. Phys.}\ }\textbf {\bibinfo {volume} {62}},\ \bibinfo
  {pages} {531} (\bibinfo {year} {1990})}\BibitemShut {NoStop}%
\bibitem [{\citenamefont {Hayes}\ \emph {et~al.}(2010)\citenamefont {Hayes},
  \citenamefont {Matsukevich}, \citenamefont {Maunz}, \citenamefont {Hucul},
  \citenamefont {Quraishi}, \citenamefont {Olmschenk}, \citenamefont
  {Campbell}, \citenamefont {Mizrahi}, \citenamefont {Senko},\ and\
  \citenamefont {Monroe}}]{Hayes2010}%
  \BibitemOpen
  \bibfield  {author} {\bibinfo {author} {\bibfnamefont {D.}~\bibnamefont
  {Hayes}}, \bibinfo {author} {\bibfnamefont {D.~N.}\ \bibnamefont
  {Matsukevich}}, \bibinfo {author} {\bibfnamefont {P.}~\bibnamefont {Maunz}},
  \bibinfo {author} {\bibfnamefont {D.}~\bibnamefont {Hucul}}, \bibinfo
  {author} {\bibfnamefont {Q.}~\bibnamefont {Quraishi}}, \bibinfo {author}
  {\bibfnamefont {S.}~\bibnamefont {Olmschenk}}, \bibinfo {author}
  {\bibfnamefont {W.}~\bibnamefont {Campbell}}, \bibinfo {author}
  {\bibfnamefont {J.}~\bibnamefont {Mizrahi}}, \bibinfo {author} {\bibfnamefont
  {C.}~\bibnamefont {Senko}},\ and\ \bibinfo {author} {\bibfnamefont
  {C.}~\bibnamefont {Monroe}},\ }\bibfield  {title} {\bibinfo {title}
  {Entanglement of atomic qubits using an optical frequency comb},\ }\href
  {https://doi.org/10.1103/PhysRevLett.104.140501} {\bibfield  {journal}
  {\bibinfo  {journal} {Phys. Rev. Lett.}\ }\textbf {\bibinfo {volume} {104}},\
  \bibinfo {pages} {140501} (\bibinfo {year} {2010})}\BibitemShut {NoStop}%
\bibitem [{\citenamefont {Lechner}\ \emph {et~al.}(2016)\citenamefont
  {Lechner}, \citenamefont {Maier}, \citenamefont {Hempel}, \citenamefont
  {Jurcevic}, \citenamefont {Lanyon}, \citenamefont {Monz}, \citenamefont
  {Brownnutt}, \citenamefont {Blatt},\ and\ \citenamefont
  {Roos}}]{Lechner2016}%
  \BibitemOpen
  \bibfield  {author} {\bibinfo {author} {\bibfnamefont {R.}~\bibnamefont
  {Lechner}}, \bibinfo {author} {\bibfnamefont {C.}~\bibnamefont {Maier}},
  \bibinfo {author} {\bibfnamefont {C.}~\bibnamefont {Hempel}}, \bibinfo
  {author} {\bibfnamefont {P.}~\bibnamefont {Jurcevic}}, \bibinfo {author}
  {\bibfnamefont {B.~P.}\ \bibnamefont {Lanyon}}, \bibinfo {author}
  {\bibfnamefont {T.}~\bibnamefont {Monz}}, \bibinfo {author} {\bibfnamefont
  {M.}~\bibnamefont {Brownnutt}}, \bibinfo {author} {\bibfnamefont
  {R.}~\bibnamefont {Blatt}},\ and\ \bibinfo {author} {\bibfnamefont {C.~F.}\
  \bibnamefont {Roos}},\ }\bibfield  {title} {\bibinfo {title}
  {Electromagnetically-induced-transparency ground-state cooling of long ion
  strings},\ }\href {https://doi.org/10.1103/PhysRevA.93.053401} {\bibfield
  {journal} {\bibinfo  {journal} {Phys. Rev. A}\ }\textbf {\bibinfo {volume}
  {93}},\ \bibinfo {pages} {053401} (\bibinfo {year} {2016})}\BibitemShut
  {NoStop}%
\bibitem [{\citenamefont {Roos}\ \emph {et~al.}(1999)\citenamefont {Roos},
  \citenamefont {Zeiger}, \citenamefont {Rohde}, \citenamefont {N\"agerl},
  \citenamefont {Eschner}, \citenamefont {Leibfried}, \citenamefont
  {Schmidt-Kaler},\ and\ \citenamefont {Blatt}}]{Roos1999}%
  \BibitemOpen
  \bibfield  {author} {\bibinfo {author} {\bibfnamefont {C.}~\bibnamefont
  {Roos}}, \bibinfo {author} {\bibfnamefont {T.}~\bibnamefont {Zeiger}},
  \bibinfo {author} {\bibfnamefont {H.}~\bibnamefont {Rohde}}, \bibinfo
  {author} {\bibfnamefont {H.~C.}\ \bibnamefont {N\"agerl}}, \bibinfo {author}
  {\bibfnamefont {J.}~\bibnamefont {Eschner}}, \bibinfo {author} {\bibfnamefont
  {D.}~\bibnamefont {Leibfried}}, \bibinfo {author} {\bibfnamefont
  {F.}~\bibnamefont {Schmidt-Kaler}},\ and\ \bibinfo {author} {\bibfnamefont
  {R.}~\bibnamefont {Blatt}},\ }\bibfield  {title} {\bibinfo {title} {Quantum
  state engineering on an optical transition and decoherence in a paul trap},\
  }\href {https://doi.org/10.1103/PhysRevLett.83.4713} {\bibfield  {journal}
  {\bibinfo  {journal} {Phys. Rev. Lett.}\ }\textbf {\bibinfo {volume} {83}},\
  \bibinfo {pages} {4713} (\bibinfo {year} {1999})}\BibitemShut {NoStop}%
\bibitem [{\citenamefont {Nagali}\ \emph {et~al.}(2012)\citenamefont {Nagali},
  \citenamefont {D'Ambrosio}, \citenamefont {Sciarrino},\ and\ \citenamefont
  {Cabello}}]{NDSC12}%
  \BibitemOpen
  \bibfield  {author} {\bibinfo {author} {\bibfnamefont {E.}~\bibnamefont
  {Nagali}}, \bibinfo {author} {\bibfnamefont {V.}~\bibnamefont {D'Ambrosio}},
  \bibinfo {author} {\bibfnamefont {F.}~\bibnamefont {Sciarrino}},\ and\
  \bibinfo {author} {\bibfnamefont {A.}~\bibnamefont {Cabello}},\ }\bibfield
  {title} {\bibinfo {title} {Experimental observation of impossible-to-beat
  quantum advantage on a hybrid photonic system},\ }\href
  {https://doi.org/10.1103/PhysRevLett.108.090501} {\bibfield  {journal}
  {\bibinfo  {journal} {Phys. Rev. Lett.}\ }\textbf {\bibinfo {volume} {108}},\
  \bibinfo {pages} {090501} (\bibinfo {year} {2012})}\BibitemShut {NoStop}%
\bibitem [{\citenamefont {Kujala}\ \emph {et~al.}(2015)\citenamefont {Kujala},
  \citenamefont {Dzhafarov},\ and\ \citenamefont {Larsson}}]{Kujala2015}%
  \BibitemOpen
  \bibfield  {author} {\bibinfo {author} {\bibfnamefont {J.~V.}\ \bibnamefont
  {Kujala}}, \bibinfo {author} {\bibfnamefont {E.~N.}\ \bibnamefont
  {Dzhafarov}},\ and\ \bibinfo {author} {\bibfnamefont {J.-A.}\ \bibnamefont
  {Larsson}},\ }\bibfield  {title} {\bibinfo {title} {Necessary and sufficient
  conditions for an extended noncontextuality in a broad class of quantum
  mechanical systems},\ }\href {https://doi.org/10.1103/PhysRevLett.115.150401}
  {\bibfield  {journal} {\bibinfo  {journal} {Phys. Rev. Lett.}\ }\textbf
  {\bibinfo {volume} {115}},\ \bibinfo {pages} {150401} (\bibinfo {year}
  {2015})}\BibitemShut {NoStop}%
\bibitem [{\citenamefont {Cirel'son}(1980)}]{cirel1980quantum}%
  \BibitemOpen
  \bibfield  {author} {\bibinfo {author} {\bibfnamefont {B.~S.}\ \bibnamefont
  {Cirel'son}},\ }\bibfield  {title} {\bibinfo {title} {Quantum generalizations
  of bell's inequality},\ }\href@noop {} {\bibfield  {journal} {\bibinfo
  {journal} {Lett. Math. Phys.}\ }\textbf {\bibinfo {volume} {4}},\ \bibinfo
  {pages} {93} (\bibinfo {year} {1980})}\BibitemShut {NoStop}%
\bibitem [{\citenamefont {Larsson}(2014)}]{larsson2014loopholes}%
  \BibitemOpen
  \bibfield  {author} {\bibinfo {author} {\bibfnamefont {J.-{\AA}.}\
  \bibnamefont {Larsson}},\ }\bibfield  {title} {\bibinfo {title} {Loopholes in
  bell inequality tests of local realism},\ }\href@noop {} {\bibfield
  {journal} {\bibinfo  {journal} {J. Phys. A: Math. Theor.}\ }\textbf {\bibinfo
  {volume} {47}},\ \bibinfo {pages} {424003} (\bibinfo {year}
  {2014})}\BibitemShut {NoStop}%
\bibitem [{\citenamefont {Borges}\ \emph {et~al.}(2014)\citenamefont {Borges},
  \citenamefont {Carvalho}, \citenamefont {de~Assis}, \citenamefont {Ferraz},
  \citenamefont {Ara\'ujo}, \citenamefont {Cabello}, \citenamefont {Cunha},\
  and\ \citenamefont {P\'adua}}]{Borges&Cabello2014}%
  \BibitemOpen
  \bibfield  {author} {\bibinfo {author} {\bibfnamefont {G.}~\bibnamefont
  {Borges}}, \bibinfo {author} {\bibfnamefont {M.}~\bibnamefont {Carvalho}},
  \bibinfo {author} {\bibfnamefont {P.-L.}\ \bibnamefont {de~Assis}}, \bibinfo
  {author} {\bibfnamefont {J.}~\bibnamefont {Ferraz}}, \bibinfo {author}
  {\bibfnamefont {M.}~\bibnamefont {Ara\'ujo}}, \bibinfo {author}
  {\bibfnamefont {A.}~\bibnamefont {Cabello}}, \bibinfo {author} {\bibfnamefont
  {M.~T.}\ \bibnamefont {Cunha}},\ and\ \bibinfo {author} {\bibfnamefont
  {S.~a.}\ \bibnamefont {P\'adua}},\ }\bibfield  {title} {\bibinfo {title}
  {Quantum contextuality in a young-type interference experiment},\ }\href
  {https://doi.org/10.1103/PhysRevA.89.052106} {\bibfield  {journal} {\bibinfo
  {journal} {Phys. Rev. A}\ }\textbf {\bibinfo {volume} {89}},\ \bibinfo
  {pages} {052106} (\bibinfo {year} {2014})}\BibitemShut {NoStop}%
\bibitem [{\citenamefont {Barz}\ \emph {et~al.}(2012)\citenamefont {Barz},
  \citenamefont {Kashefi}, \citenamefont {Broadbent}, \citenamefont
  {Fitzsimons}, \citenamefont {Zeilinger},\ and\ \citenamefont
  {Walther}}]{barz2012demonstration}%
  \BibitemOpen
  \bibfield  {author} {\bibinfo {author} {\bibfnamefont {S.}~\bibnamefont
  {Barz}}, \bibinfo {author} {\bibfnamefont {E.}~\bibnamefont {Kashefi}},
  \bibinfo {author} {\bibfnamefont {A.}~\bibnamefont {Broadbent}}, \bibinfo
  {author} {\bibfnamefont {J.~F.}\ \bibnamefont {Fitzsimons}}, \bibinfo
  {author} {\bibfnamefont {A.}~\bibnamefont {Zeilinger}},\ and\ \bibinfo
  {author} {\bibfnamefont {P.}~\bibnamefont {Walther}},\ }\bibfield  {title}
  {\bibinfo {title} {Demonstration of blind quantum computing},\ }\href
  {https://doi.org/10.1126/science.1214707} {\bibfield  {journal} {\bibinfo
  {journal} {Science}\ }\textbf {\bibinfo {volume} {335}},\ \bibinfo {pages}
  {303} (\bibinfo {year} {2012})}\BibitemShut {NoStop}%
\bibitem [{\citenamefont {Reichardt}\ \emph {et~al.}(2013)\citenamefont
  {Reichardt}, \citenamefont {Unger},\ and\ \citenamefont
  {Vazirani}}]{reichardt2013classical}%
  \BibitemOpen
  \bibfield  {author} {\bibinfo {author} {\bibfnamefont {B.~W.}\ \bibnamefont
  {Reichardt}}, \bibinfo {author} {\bibfnamefont {F.}~\bibnamefont {Unger}},\
  and\ \bibinfo {author} {\bibfnamefont {U.}~\bibnamefont {Vazirani}},\
  }\bibfield  {title} {\bibinfo {title} {Classical command of quantum
  systems},\ }\href@noop {} {\bibfield  {journal} {\bibinfo  {journal}
  {Nature}\ }\textbf {\bibinfo {volume} {496}},\ \bibinfo {pages} {456}
  (\bibinfo {year} {2013})}\BibitemShut {NoStop}%
\bibitem [{\citenamefont {Bharti}\ \emph {et~al.}(2019)\citenamefont {Bharti},
  \citenamefont {Ray}, \citenamefont {Varvitsiotis}, \citenamefont {Warsi},
  \citenamefont {Cabello},\ and\ \citenamefont {Kwek}}]{Bharti2019}%
  \BibitemOpen
  \bibfield  {author} {\bibinfo {author} {\bibfnamefont {K.}~\bibnamefont
  {Bharti}}, \bibinfo {author} {\bibfnamefont {M.}~\bibnamefont {Ray}},
  \bibinfo {author} {\bibfnamefont {A.}~\bibnamefont {Varvitsiotis}}, \bibinfo
  {author} {\bibfnamefont {N.~A.}\ \bibnamefont {Warsi}}, \bibinfo {author}
  {\bibfnamefont {A.}~\bibnamefont {Cabello}},\ and\ \bibinfo {author}
  {\bibfnamefont {L.-C.}\ \bibnamefont {Kwek}},\ }\bibfield  {title} {\bibinfo
  {title} {Robust self-testing of quantum systems via noncontextuality
  inequalities},\ }\href {https://doi.org/10.1103/PhysRevLett.122.250403}
  {\bibfield  {journal} {\bibinfo  {journal} {Phys. Rev. Lett.}\ }\textbf
  {\bibinfo {volume} {122}},\ \bibinfo {pages} {250403} (\bibinfo {year}
  {2019})}\BibitemShut {NoStop}%
\bibitem [{\citenamefont {Lunghi}\ \emph {et~al.}(2015)\citenamefont {Lunghi},
  \citenamefont {Brask}, \citenamefont {Lim}, \citenamefont {Lavigne},
  \citenamefont {Bowles}, \citenamefont {Martin}, \citenamefont {Zbinden},\
  and\ \citenamefont {Brunner}}]{Lunghi2015}%
  \BibitemOpen
  \bibfield  {author} {\bibinfo {author} {\bibfnamefont {T.}~\bibnamefont
  {Lunghi}}, \bibinfo {author} {\bibfnamefont {J.~B.}\ \bibnamefont {Brask}},
  \bibinfo {author} {\bibfnamefont {C.~C.~W.}\ \bibnamefont {Lim}}, \bibinfo
  {author} {\bibfnamefont {Q.}~\bibnamefont {Lavigne}}, \bibinfo {author}
  {\bibfnamefont {J.}~\bibnamefont {Bowles}}, \bibinfo {author} {\bibfnamefont
  {A.}~\bibnamefont {Martin}}, \bibinfo {author} {\bibfnamefont
  {H.}~\bibnamefont {Zbinden}},\ and\ \bibinfo {author} {\bibfnamefont
  {N.}~\bibnamefont {Brunner}},\ }\bibfield  {title} {\bibinfo {title}
  {Self-testing quantum random number generator},\ }\href
  {https://doi.org/10.1103/PhysRevLett.114.150501} {\bibfield  {journal}
  {\bibinfo  {journal} {Phys. Rev. Lett.}\ }\textbf {\bibinfo {volume} {114}},\
  \bibinfo {pages} {150501} (\bibinfo {year} {2015})}\BibitemShut {NoStop}%
\end{thebibliography}%

\end{document}